\documentclass[useAMS,fleqn,usenatbib,a4paper]{mn2e} 

\usepackage{times}  
\usepackage{graphicx}

\usepackage{amsmath,amssymb}
\usepackage{bm} 
\usepackage{fixltx2e} 

\renewcommand{\vec}[1]{\ensuremath{\bm{#1}}} 

\makeatletter
\providecommand*{\diff}{\@ifnextchar^{\DIfF}{\DIfF^{}}}
\def\DIfF^#1{\mathop{\mathrm{\mathstrut d}}\nolimits^{#1}\gobblespace}
\def\gobblespace{\futurelet\diffarg\opspace}
\def\opspace{\let\DiffSpace\!\ifx\diffarg(\let\DiffSpace\relax
                                                  \else\ifx\diffarg[\let\DiffSpace\relax
                                                  \else\ifx\diffarg\{\let\DiffSpace\relax\fi\fi\fi\DiffSpace}
\providecommand*{\degree}{\ensuremath{^{\circ}}}
\providecommand*{\unit}[1]{\ensuremath{\mathrm{\,#1}}}
\newcommand{\kms}{\ensuremath{\unit{km}\unit{s}^{-1}}}

\newcommand{\ualpha}{\alpha}
\newcommand{\ubeta}{\beta}
\newcommand{\ugamma}{\gamma}

\newcommand{\boldtxt}[1]{#1}

\title[CSM in RS Oph]{Modelling the circumstellar medium in RS Ophiuchi and its link to Type Ia supernovae}

\author[Booth, Mohamed \& Podsiadlowski]
	{R.~A.~Booth,$^{1,2}$\thanks{E-mail: rab200@ast.cam.ac.uk} S.~Mohamed$^3$ and Ph.~Podsiadlowski$^2$\\
        $^1$Institute of Astronomy, University of Cambridge, Madingley Road
                 Cambridge, CB3 0HA, UK\\
	$^2$Department of Astrophysics, University of Oxford,
		 Keble Road, Oxford OX1 3RH, UK\\
	$^3$South African Astronomical Observatory, PO Box 9,
		Observatory 7935, Cape Town, Western Cape, South Africa}

\pagerange{\pageref{firstpage}--\pageref{lastpage}} \pubyear{2015}

\begin{document}

\label{firstpage}

\maketitle

\begin{abstract}
Recent interpretations of narrow, variable absorption lines detected in some Type Ia supernovae suggest that their progenitors are surrounded by dense, circumstellar material. Similar variations detected in the symbiotic recurrent nova system RS Oph, which undergoes thermonuclear outbursts every ~20 years, making it an ideal candidate to investigate the origin of these lines. To this end, we present simulations of multiple mass transfer-nova cycles in RS Oph. We find that the quiescent mass transfer produces a dense, equatorial outflow, i.e., concentrated towards the binary orbital plane, and an accretion disc forms around the white dwarf. The interaction of a spherical nova outburst with these aspherical circumstellar structures produces a bipolar outflow, similar to that seen in HST imaging of the 2006 outburst. In order to produce an ionization structure that is consistent with observations, a mass-loss rate of $5 \times 10^{-7}\unit{M}_{\sun}\unit{yr}^{-1}$ from the red giant is required. The simulations also produce a polar accretion flow, which may explain the broad wings of the quiescent H α line and hard X-rays. By comparing simulated absorption line profiles to observations of the 2006 outburst, we are able to determine which components arise in the wind and which are due to the novae. We explore the possible behaviour of absorption line profiles as they may appear should a supernova occur in a system like RS Oph. Our models show similarities to supernovae like SN 2006X, but require a high mass-loss rate, $\dot{M} \sim 10^{-6}$ to $10^{-5}\unit{M_\odot}\unit{yr}^{-1}$, to explain the variability in SN 2006X.

\end{abstract}


\section{Introduction}
RS Ophiuchi (RS Oph) is a symbiotic recurrent nova; the system consists of a red giant and a white-dwarf companion that undergoes nova outbursts approximately every 20 years. The most recent outburst was in 2006,  in which a sustained multi-wavelength campaign resulted in improved estimates of the explosion parameters and geometry. Resolved VLBI and \textit{HST} images confirmed the aspherical nature of the nova shell \citep{O'Brien2006,Bode2007,Rupen2008}, and early X-ray emission from the outburst showed evidence of interaction between the nova and a dense circumstellar medium (CSM) \citep{Sokoloski2006,Vaytet2011}. The outburst was characterised by low ejecta mass, $M_{\rm ej} \sim 10^{-7}$ to  $10^{-6}\,\unit{M}_{\sun}$ \citep{Sokoloski2006}, and high ejecta velocity, $v_{\rm ej}  \sim 4000\kms$ \citep{Das2006,Anupama2008}.  The ejecta mass and velocity estimates, together with the fast decline of the nova \citep{Hounsell2010}, indicate that the white dwarf is close to the Chandrasekhar mass \citep{Yaron2005}. Combined with the fact that the white dwarf mass may be increasing \citep{Hachisu2007,Hernanz2008}, this makes RS Oph an ideal candidate Type Ia supernova progenitor. 

A further possible connection between RS Oph and Type Ia supernovae (SNe Ia) is the detection of narrow, time-variable absorption lines, which may be related to the presence of a circumstellar medium around the supernova progenitor. Time-variable absorption lines were first detected in  the Type Ia supernova SN 2006X  \citep{Patat2007}, in which the Na\,\textsc{i} D lines were first observed to strengthen around maximum light, and then weaken by 60 days after the explosion. This behaviour was suggested to originate in surrounding nova shells, which first recombine and subsequently are swept up by the supernova. The RS Oph -- supernova connection is further supported by the 2006 outburst of RS Oph, in which similar behaviours were seen in the Na\,\textsc{i} D and Ca\,\textsc{ii} H \& K lines \citep{Patat2011}. Time-variable absorption has since been detected in several other SNe Ia \citep{Simon2009,Blondin2009,Stritzinger2010} and statistical studies suggest that 20 to 25 per cent of SNe Ia may have pre-supernova outflows \citep{Sternberg2011,Maguire2013}. The origin of these lines has been the subject of a range of theoretical studies, from red giant winds and recurrent novae \citep{Chugai2008,Moore2012} to tidal tails ejected from double-degenerate systems \citep{Raskin2013} or even novae in He + CO systems \citep{Shen2013}. 

Spectroscopic measurements of the velocities of both stars has enabled an accurate determination of the orbital period, 453.6 days, mass ratio, $q = 0.6$ and an eccentricity, $e=0$ \citep{Dobrzycka1994,Shore1996,Fekel2000,Brandi2009}. Combined with the estimate for the white-dwarf mass, this allows an accurate determination of the red-giant (RG) mass,  $M_\mathrm{RG} \approx 0.8 \unit{M}_{\sun}$, inclination, $i \approx 50\degree{}$ and separation, $a \approx 1.48\unit{AU}$. Measurements of the rotation period of the giant suggest that the giant is close to filling its Roche lobe unless the red giant is super-synchronous, i.e. the rotation period of the giant is less than that of the binary \citep{Zamanov2007,Brandi2009}. Such tight constraints on the physical parameters make RS Oph an ideal candidate for testing the recurrent nova model of SNe Ia.

Observations provide a number of constraints on the CSM in RS Oph. Direct estimates of the quiescent mass-loss rates are somewhat uncertain, varying from between $ \dot{M} \approx 10^{-8}\,\unit{M}_{\sun} \unit{yr}^{-1}$, as estimated by dust emission \citep{Evans2007}, to  $ \dot{M} \approx 10^{-6}\,\unit{M}_{\sun} \unit{yr}^{-1}$ from Na\,\textsc{i} D absorption lines during the 2006 outburst \citep{Iijima2008}. The Na\,\textsc{i} D lines provide a measure of the wind velocity, from which \citet{Iijima2008} obtained a velocity of $33\kms$. High-resolution spectroscopy shows a range  blue-shifted (out-flowing) components from 10 to $37\kms$. While the slower components were weaker after the nova, the fastest components strengthened \citep{Patat2011}. The complex structure of the wind is further demonstrated by the presence at all epochs of a red-shifted (in-flowing) absorption component with a velocity  of $v \lesssim 20\kms$. 

Understanding how and where the absorption lines form in the CSM is key to understanding whether recurrent novae are responsible for the origin of the lines in SN Ia. Simulations have already shown strong asymmetry in the mass loss from binary systems \citep{Theuns1993,Mastrodemos1998,Mohamed2007,Walder2008}, which needs to be taken into account in recurrent nova models and their relation to SNe Ia.

In section \ref{Sec:CSMModel} we present simulations of mass transfer in RS Oph. We include multiple novae and build a model of the CSM around RS Oph. Using photoionization calculations we constrain the quiescent mass-loss rate in RS Oph.  In section \ref{Sec:Na} we calculate theoretical line profiles before and after the novae. By comparing them to observations, we determine the origin of these components in RS Oph. In section \ref{Sec:Ha} we investigate a polar inflow on to the white dwarf as a possible explanation for the broad wings observed in hydrogen lines during quiescence. Finally, we discuss the results in terms of the prospects for SNe Ia in section \ref{Sec:Discussion}.

\section{Circumstellar Model}
\label{Sec:CSMModel}
\subsection{Quiescent Mass-loss Phase}
\label{Sec:Quies}

The RS Oph binary system, quiescent mass loss and nova outbursts were modelled using the Smoothed Particle Hydrodynamics (SPH) code \textsc{gadget-2} \citep{Springel2005}, which has been modified to include binary motion, stellar winds, accretion and cooling \citep{Mohamed2010, Mohamed2012}. For temperatures above $10^4\unit{K}$, the NEI $[\mathrm{Fe/H}] = -0.5$ cooling curves from \citet{Sutherland1993} were used. Following \citet{Mohamed2012}, fine-structure and molecular cooling processes including H$_2$, CO and H$_2$O were used at temperatures below $10^4\unit{K}$.

For the binary parameters, we assumed the values derived from the spectroscopic orbit for RS Oph \citep{Brandi2009} in which the binary orbital period was 453.6 days. The internal structures of the white dwarf and red giant were not modelled in detail, instead their gravity was treated assuming point potentials with masses of $1.38\unit{M}_{\sun}$ and $0.8\unit{M}_{\sun}$, respectively. The  mass loss from the red giant {was} modelled by injecting the particles close to the surface to give a mass-loss rate of approximately $10^{-7}\unit{M}_{\sun}\unit{yr}^{-1}$. The particles were injected with a low velocity, $v = 20\unit{km}\unit{s}^{-1}$, much less than the escape speed of the RG, $v_\mathrm{esc} \approx 55 \unit{km}\unit{s}^{-1}$. \boldtxt{While this is sufficient to lift the material beyond the red giant's Roche lobe, the material remains bound to the binary. The escape velocity is finally achieved via continued tidal acceleration of circumbinary material by the binary.}

\boldtxt{Since our simulations are aimed at studying the large scale structure of the circumstellar environment of recurrent novae, it is not feasible to also simulate in detail the flow structure close to the white dwarf due to the very different time-scales involved (tens to hundreds of years for the novae and hours to days for the flow accreting onto the white dwarf). For this reason we use a simple prescription to treat accretion onto the white dwarf, and do not include any explicit mass-loss from the white dwarf or its accretion disc, and treat the white dwarf as a sink. Following \citet{Theuns1993} and \citet{Mohamed2012b}, the mass of particles within $\tilde{h}$ of the white dwarf is decreased by a factor $(r/\tilde{h})^2$ in each time-step, where $\tilde{h}$ is the average smoothing length of particles within $0.1\unit{au}$ of the white dwarf. Particles are removed once their mass drops below one per cent of their initial mass. This effectively ensures no pressure build up in the sink region.}

\boldtxt{While this undoubtedly affects the structure of the flow close to the white dwarf, we have found that it makes very little difference to the large scale structure of the wind (excepting the cooling and temperature of the wind, discussed below). Despite the uncertainties associated with our treatment, the flow close to the binary has some interesting features that may be able to explain a number of interesting observational properties of RS Oph. Therefore, we take the approach of describing these features, along with the appropriate caveats and further possible evidence for these features.}

\begin{figure*}
\centering
\includegraphics[width=\textwidth]{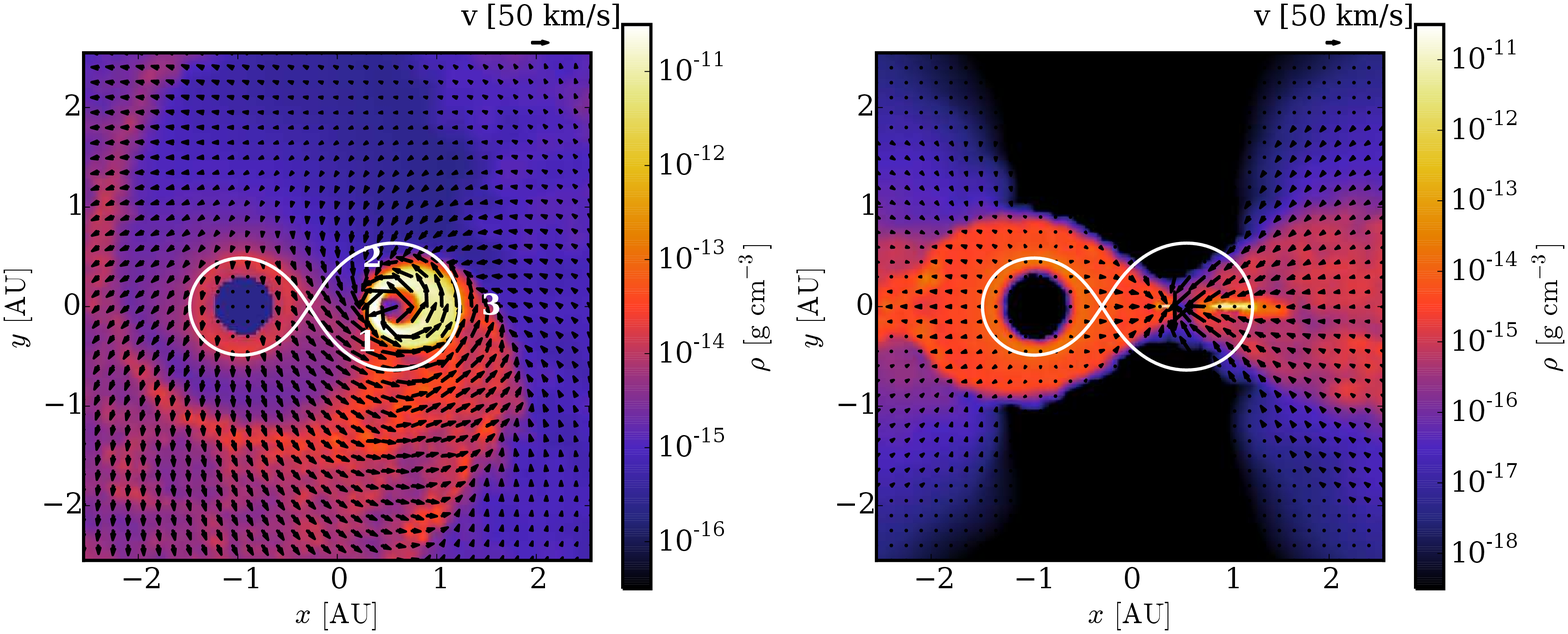}
\vspace{0 mm}

\includegraphics[width=\textwidth]{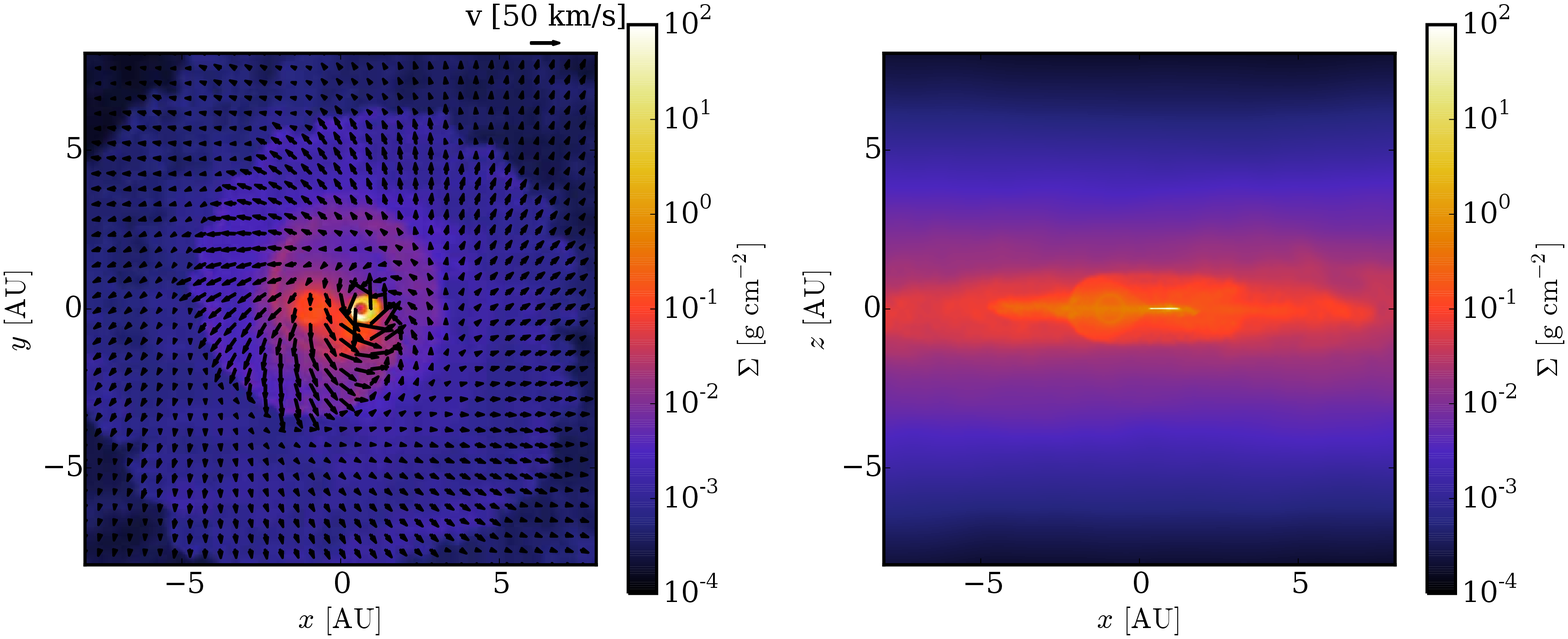}
\caption{Spiral structure formed during the RS Oph quiescent mass-loss phase. The red giant is to the left of the white dwarf and accretion disc. \emph{Upper Panel}:  Density cross-sections through the flow structure and the velocity relative to the centre of mass. Part of the wind falls back onto the white dwarf from above and below the plane of the binary. \boldtxt{The numbers show the Roche lobe overflow stream (1), which interacts with the wind swept up during the orbit (2) to form a spiral wind, and the wind escaping through the L1 Lagrange point (3), which forms another spiral. The solid white lines show the sizes of the Roche lobe for the red giant and white dwarf.} \emph{Lower Panel}:  Large-scale structure, showing the merger of the two spiral streams and the stratification of the wind. \boldtxt{The colour bar shows column density integrated over $z/y$. At distances greater than a few orbital radii, the wind becomes ballistic and the velocity is approximately radial and perpendicular to the spirals.}
}
\label{QuiesStruct}
\end{figure*}

The circumstellar wind structure formed after 13 orbits is shown in Fig.~\ref{QuiesStruct}. \boldtxt{The structure formed by the interaction of a slow wind with a binary companion has been discussed in detail by \citet{Theuns1993}. We therefore only repeat the details needed in the present work.} The binary potential confines the wind to binary orbital plane, forming a two-armed spiral structure. One spiral forms in the flow passing the white dwarf and escaping through one of the outer Lagrange points \boldtxt{(see Fig.~\ref{QuiesStruct}, label 3)}. The second forms due to the interaction of the Roche lobe overflow stream that passes through the inner Lagrange point, with the part of the wind passing the white dwarf \boldtxt{(flows labelled 1 and 2, Fig.~\ref{QuiesStruct})} . These flows can be seen in the top left panel of Fig.~\ref{QuiesStruct}, with the Roche lobe overflow stream flowing from left to right, and the wind arriving from the top.

The two spirals run into each other, forming a shock. The streams merge after approximately 1.25 orbits. Behind the shock, the spiral becomes clumpy, driven by cooling, which is thermally unstable around $10^3\unit{K}$. The clumps form at the resolution scale. \boldtxt{(while the resolution scale is typically less than the size of structures in the flow, the physical size of the clumps is likely much smaller).} 

An accretion disc forms which  extends to \boldtxt{the edge of the white dwarf's Roche lobe}, approximately $10^{13}\unit{cm}$. The formation of the accretion disc is enabled by the cooling; similarly to \citet{Theuns1993}, we found no accretion disc forms in simulations where cooling is neglected.  The outer parts of the accretion disc are on an elliptical orbit that remains aligned along the line between the red giant and white dwarf. \boldtxt{The accretion disc is only resolved vertically by a few smoothing lengths in the simulations and since we have neglected sources of heating the thickness of the accretion disc is likely an underestimate. We also find the mass in the accretion disc is resolution dependent (higher in higher resolution simulations). The accretion rate is not sensitive to resolution since it is controlled by the mass flow into the white dwarf's Roche lobe, which suggests an explanation for the varying accretion disc mass. Since the artificial viscosity is lower in higher resolution simulations a higher disc mass is needed to maintain a similar accretion rate and therefore the disc mass adjusts itself to give the same accretion rate in steady state \boldtxt{($1\mbox{--} 2 \times 10^{-8}\,M_\odot\unit{yr}^{-1}$)}. Due to the uncertainty in the accretion disc properties from the simulations, the discussion will be guided by observations where possible.}

\boldtxt{The passage of the white dwarf through the wind of the red giant as the binary rotates gives rise to a component of the wind that falls onto the white dwarf and accretion disc, as was originally seen in the simulations of \citet{Theuns1993}. While the details of this flow are dependent on the structure of the accretion disc, the presence of \emph{some} material falling onto the white dwarf and accretion disc reflects the fact that the pressure scale height in the accretion disc is smaller than the scale height in the part of the wind that is close the binary. Therefore the wind that passes above the white dwarf is unable to support itself and falls onto the white dwarf and accretion disc. The polar stream that arises in our simulations contributes an accretion rate onto the white dwarf of order ${\dot{M} \sim 5 \times  10^{-9}\unit{M}_{\sun}\unit{yr}^{-1}}$. In sections~\ref{Sec:Ha} and \ref{Sec:Na} we discuss the evidence for a polar accretion component.}

\begin{figure}
\centering
\includegraphics[width=\columnwidth]{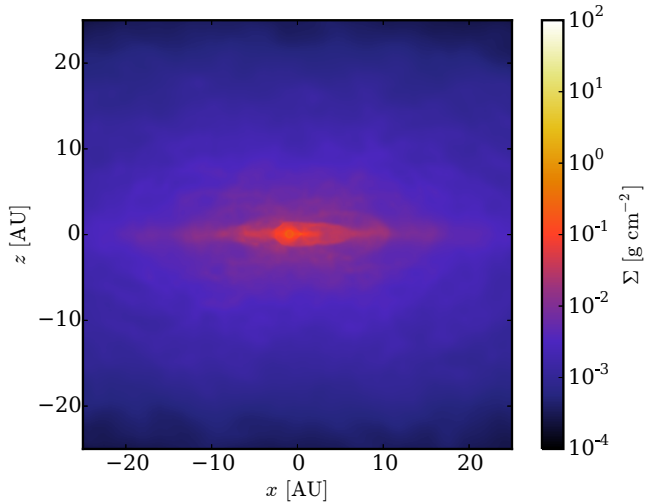}
\caption{\boldtxt{Column density in a model with no cooling, which produces a less strongly confined wind structure. For comparison, the colour bar shows the same column density range as Fig.~\ref{QuiesStruct}, but the spatial scale is 2.5 times larger to demonstrate the larger vertical extent of the flow.}
}
\label{AdiabatStruct}
\end{figure}

\boldtxt{Although the wind becomes ballistic beyond a few binary separations, the temperature of the wind close to the binary is important for controlling the vertical scale height in the wind and depends on the heating and cooling models used. While cooling has been included in the models presented, no contribution to heating from the white dwarf, accretion disc or red giant has been included. This results in low temperatures in the wind close to the binary and  a wind that is highly confined to the binary plane. In models in which the cooling has been suppressed the wind is less strongly confined to binary plane, resulting in lower densities near the plane but higher densities at higher latitudes, although clear asymmetries are still present. This is demonstrated in Fig.~\ref{AdiabatStruct}, which shows a model in which the flow is adiabatic. Even in this extreme case, the wind remains asymmetric. The wind in RS Oph is likely to lie between these two extremes --} the effect this has on relationship to observations of RS Oph is discussed further in section \ref{Sec:Na}.

\subsection{Novae Outbursts}

\begin{figure*}
\centering
\includegraphics[width=\linewidth]{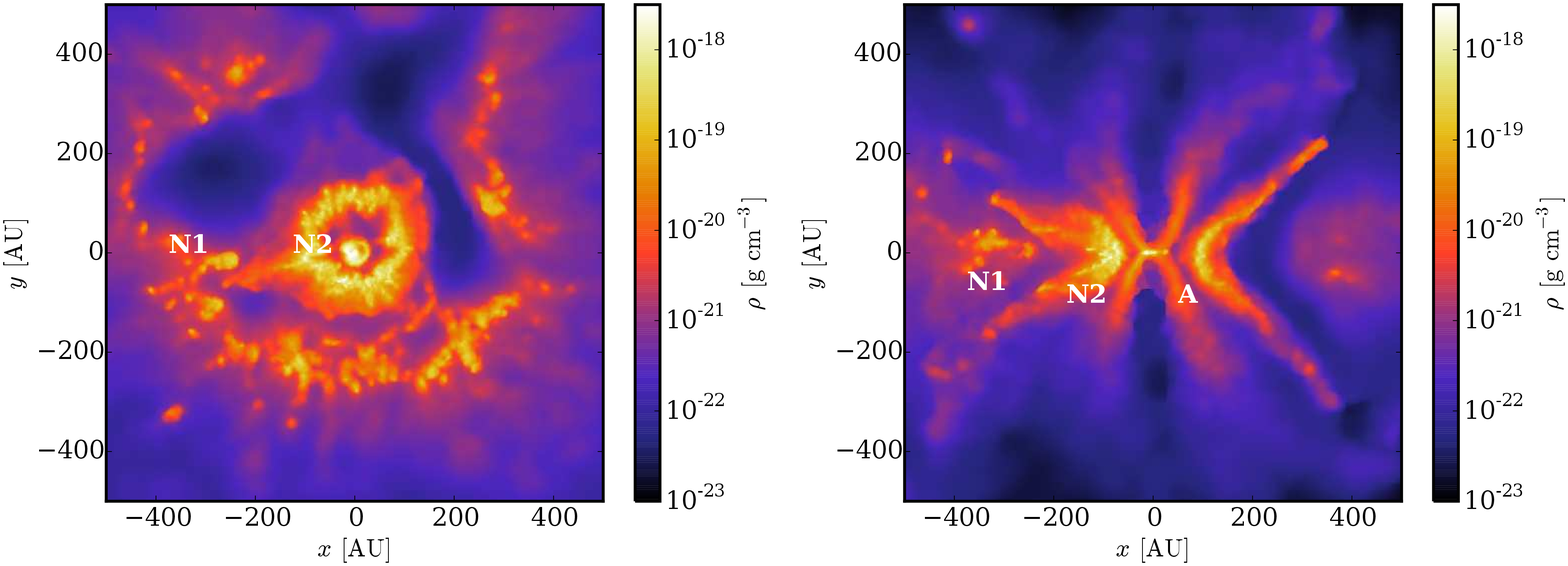}
\vspace{0 mm}

\includegraphics[width=\textwidth]{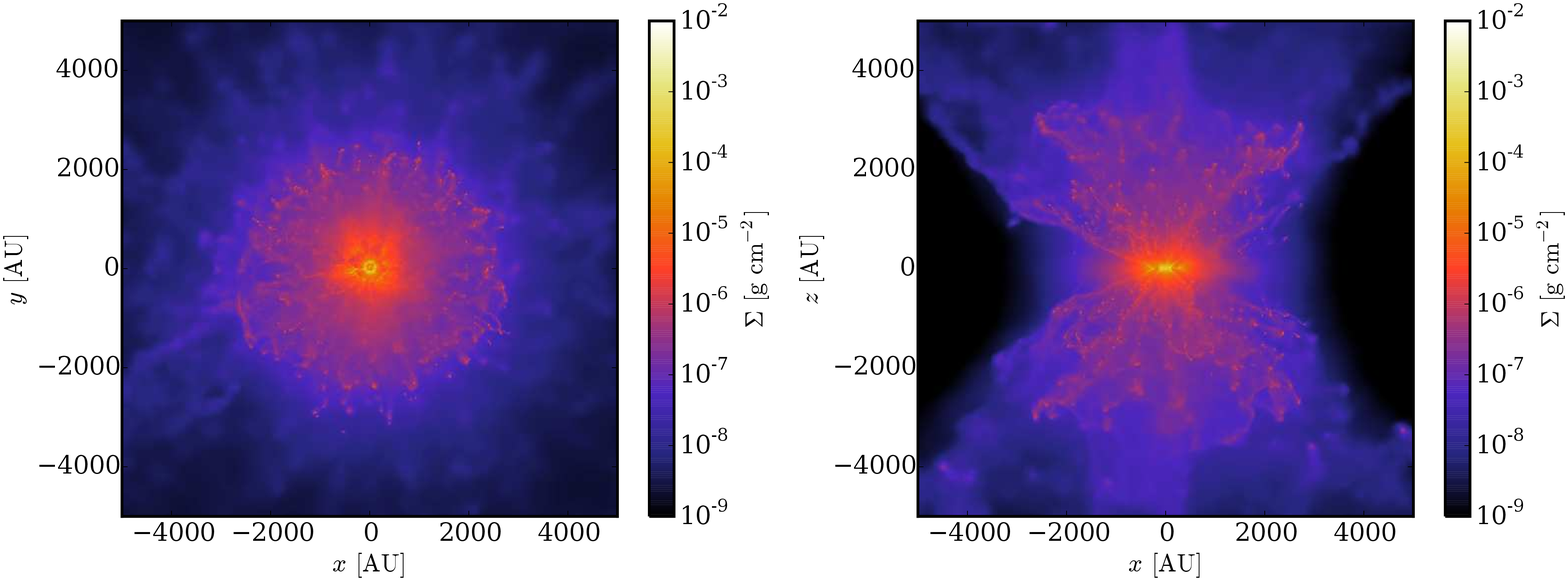}
\caption{Density structure 2000 days after the second nova outburst. \emph{Upper Panel}:  Zoom-in of the equatorial ring. \boldtxt{The labels show the location of the equatorial rings from the first (N1) and second novae (N2), along with the `evaporative' wind from the accretion disc (A). The first nova shell (N1) is more clumpy, reflecting the additional cooling phase that it underwent after being shocked by the ejecta from the second nova.} \emph{Lower Panel}: Large scale structure of the novae, \boldtxt{showing the bipolar structure shaped by the interaction with the accretion disc and the wind. The velocity of the bipolar lobes is much higher ($4000\kms$) than the slow moving equatorial ring ($200\kms$).}}

\label{Novae}
\end{figure*}

After the quiescent mass transfer has evolved for 18 years, a nova shell was injected into the simulation. Subsequently the interaction between multiple novae and the circumstellar medium were simulated. The novae were modelled using a uniform shell for the ejecta, inserted within the sink region around the white dwarf. The nova mass was taken to be $M_{\rm ej} \approx 2 \times 10^{-7} \unit{M}_{\sun}$, and the velocity was set to produce a terminal velocity, $v_{\rm ej} \approx 3500\unit{km}\unit{s}^{-1}$. The internal energy of the ejecta was initially set to 25 per cent of the kinetic energy, ensuring the dynamics are dominated by the ram pressure. 

Three novae, each separated by 18 years, have been modelled, each using the circumstellar structure formed by the previous novae. Rather than re-compute the quiescent mass loss, we assumed the presence of the previous novae had no effect on this phase. Since the gas remaining from the previous novae is much less dense than the material lost during quiescence, any error made in the large scale wind structure should be small.

Within the first day the nova ejecta interacts with the accretion disc. Since the gas in the accretion disc is much denser than the nova ejecta, it immediately distorts the nova shell, restricting the flow in the binary orbital plane and forcing the nova shell to become bipolar. \boldtxt{The resulting structures are shown in Fig.~\ref{Novae}. The large wind mass close to the orbital plane causes the nova shells to decelerate to velocities of 100 to $200\kms$ as the entire wind during the previous quiescent phase is swept up over the course of a few years. This results in the formation of an equatorial ring-like structure (components N1 and N2, Fig.~\ref{Novae}) that is considerably more dense than the rest of the nova shell. While the bipolar lobes primarily consist of nova ejecta, the equatorial ring consists mainly of swept up wind material and will therefore have a similar composition to the red giant. In the polar directions, the nova shell rapidly escapes the CSM and remains close to its initial velocity. As the forward shock travels down the density gradient in the polar direction, it accelerates, and the swept-up circumstellar gas reaches velocities of $4000\kms$ to $4500\kms$. The high velocity of ejecta in the polar directions means that the ejecta catch up and drive shocks through the equatorial rings of the previous novae. This interaction leads to clump formation as the shell cools and further enhances the clumping in the old shells. While the repeated shocking and cooling of the older nova shells causes the shells to partially break up, multiple equatorial rings maybe visible.}

We find that the accretion disc survives the passage of the nova since it is considerably more dense than the nova ejecta. \boldtxt{This result is somewhat sensitive to our assumptions, since in some of the lower resolution simulations the accretion disc did not survive, due to the lower mass in the accretion disc at lower resolutions. The survival of the accretion disc will also depend on the details of its structure, such as its scale height. However, our simulations most likely underestimate the mass in the accretion disc and observations during the 2006 outburst of RS Oph suggest that the disc survives \citep{Hachisu2006}. The interaction of the nova ejecta with the accretion disc immediately strips the weakly bound gas from the surface of the disc, which gets entrained in the nova shell. Additionally, once the nova ejecta has passed the accretion disc (within hours in our simulations), we see evidence for an additional phase of mass loss. This mass loss occurs because in addition to stripping the gas, the nova drives shocks through the accretion disc, heating it up. The high pressure in the shocked gas then drives an additional `evaporative wind' from the surface of the accretion disc, which lasts for several days and appears as an additional shell-like feature (component A, Fig.~\ref{Novae}). A similar effect provides an important contribution to the total mass lost when a supernova interacts with a binary companion \citep{Wheeler1975}.}

The highest density in the nova shells is in the equatorial ring, which is $n_\mathrm{H} \sim 10^6\unit{cm}^{-3}$ 2000 days after the nova. This density is lower than the $10^7\unit{cm}^{-3}$ predicted by 1D analytical models \citep{Moore2012} scaled to a comparable mass-loss rate to the effective mass-loss rate in the binary plane. \boldtxt{The discrepancy may be due to a variety of reasons. One possibility is that the 3D geometry allows the shell to spread out in the vertical directions, which is impossible in 1D. Another possibility is that the simulations may not sufficiently resolve the dense shell.} This introduces an uncertainty into the recombination time-scale,  which depends on the density through $t_r = (n_e \alpha)^{-1}$, where $\alpha$ is the recombination coefficient  and $n_e$ is the electron density. Although the velocity of the absorption lines formed in the nova shells should be reliable, since it is controlled by momentum conservation, the uncertainty in the recombination time-scale means the strength of these components remains only qualitative (see Sec.~\ref{Sec:Ion}).

\subsection{Mass-loss rate estimate from the ionization state of the circumstellar medium}
\label{Sec:Ion}

\begin{figure}
\includegraphics[width=0.45\textwidth]{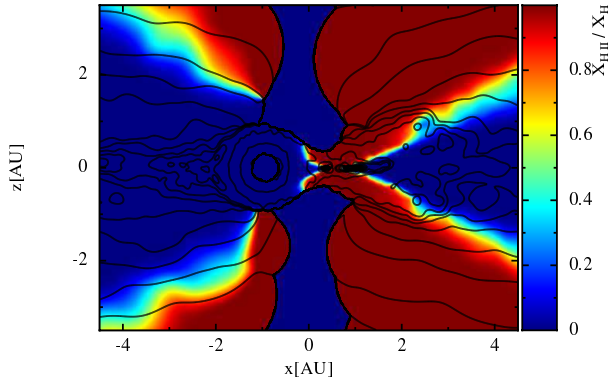}
\caption{Structure of the hydrogen ionization fraction during quiescence, for a cross-section through the red giant and white dwarf. The wind close to the red giant and the accretion disc provide sufficient optical depth to shield the gas behind them, elsewhere the gas is ionized. The three-dimensional ionization structure is therefore conical, centred on the white dwarf, with the additional neutral region behind the red giant. \boldtxt{Contours show the density. Note the spatial scale is approximately twice as wide as in Fig.~\ref{QuiesStruct}. To compare them directly, the density contours show clearly the edge of the dense wind region near the RG.}}
\label{Ioniz}
\end{figure}

The ionization conditions of stellar winds can be an effective tracer of the mass-loss rate, since, neglecting optical depth, the photoionization rate from a point source scales as $r^{-2}$, and the recombination rate scales as $r^{-4}$ in a one-dimensional wind. This means that unless the optical depth is sufficient to absorb the ionizing flux the entire wind will be ionized. In RS Oph, hydrogen is only ionized relatively close to the white dwarf \citep{Anupama1999}, an idea supported by our Na\,\textsc{i} D line profile modelling, therefore we can use this to estimate the mass-loss rate from the red giant.

During quiescence the dominant ionizing flux in RS Oph comes from the hot component, although whether this flux is dominated by the white dwarf or accretion disc is unknown. \citet{Dobrzycka1996} used UV observations to derive a hot component luminosity of $100-300\,L_\odot$, and a temperature $T \approx 10^5\unit{K}$. Although this is the luminosity after the emission has been reprocessed by the region close to the white dwarf \citep{Anupama1999}, it is the appropriate luminosity for calculating the ionization state of material in the wind. Soft X-rays emitted as a result of the novae outbursts also contribute to the ionizing flux. During quiescence an X-ray luminosity of $2 \times 10^{32} \unit{erg}\unit{s}^{-1}$ was measured by \textit{RXTE} in 1997, roughly 12 years after the 1985 outburst \citep{Mukai2008}. 

The quiescent hydrogen ionization state was modelled using a Str\"omgren sphere approximation for lines of sight from each particle to the white dwarf, with each line of sight treated independently and the ionization of each particle found by balancing the UV photoionization with Case B recombination, taking into account the optical depth along the line of sight. Case B recombination assumes that any photons emitted with $E > 13.6\unit{eV}$ are reabsorbed locally by another hydrogen atom and can be neglected from the overall recombination rate. The effective photoionization cross-section is calculated by averaging the cross-section for photoionization from the ground state over a black-body spectrum at $10^5\unit{K}$. The Case B recombination coefficient at $10^4\unit{K}$ from \citet{Ferguson1997} is used and the photoionization cross-section is from \citet{Verner1996b}.

The results of the Str\"omgren calculation are shown in Fig.~\ref{Ioniz}. Only the accretion disc and wind closest to the red giant provide sufficient optical depth that the gas becomes neutral, resulting in a conical ionization structure, with a larger neutral region in the gas shielded by the red giant. The low UV optical depth  is in disagreement with \citet{Shore1993} and \citet{Anupama1999}, who suggest a large optical depth is required to explain the line ratios, and  that the hot component is embedded in a neutral wind. The strong absorption seen in the Balmer series \citep{Bulla2013} and the presence of metal absorption lines such as the Na\,\textsc{i} D lines that form in the wind \citep{Patat2011} also support a large optical depth.

The larger density close to the white dwarf must mean a higher mass-loss rate, since the simulations over-estimate the density in the wind because they neglected photoionization heating. Recalculating the ionization structure for models in which the density has been scaled to mimic an increased mass-loss rate, we find that a mass-loss rate $\dot{M} = 5 \times 10^{-7}\unit{M}_{\sun}\unit{yr}^{-1}$ is sufficient to produce an ionized bubble around the white dwarf with a neutral region outside it. The relatively modest increase in mass-loss rate is sufficient since the optical depth is sensitive to density; in a highly ionized region, the neutral fraction $Y \approx n \alpha / F \sigma$, \boldtxt{where $F$ is the flux of ionizing photons, $n$ is the number density of hydrogen and $\sigma$ is the ionization cross-section. This results in the optical depth, $\tau$, obeying} $\tau \propto \dot{M}^2$. The optical depths for the simulated Na\,\textsc{i}\,D lines discussed in section \ref{Sec:Na} also support a higher mass-loss rate in RS Oph.

In order to calculate the Na\,\textsc{i}\,D line profiles, the ionization state of sodium must also be modelled, because its low ionization energy (5.1\unit{eV}) means sodium can remain photoionized in regions where hydrogen is neutral. A detailed, time-dependent and multi-species photoionization model in 3D is beyond the scope of this work, instead we use a model that captures the important quantities for sodium, i.e., the photoionization rate and the recombination rate, which depends on the local electron density. We also include collisional ionization  which can dominate in the novae; where the gas behind the forward shock can reach $10^6 \unit{K}$. The photoionization, recombination and collisional ionization coefficients from \citet{Verner1996a}, \citet{Verner1996b} and \citet{Arnaud1985} were used.

Since sodium is fully ionized in regions where hydrogen is ionized we focused on the ionization state of sodium in the neutral hydrogen regions. Since in this region the optical depth to UV photons is large, we modelled the ionization of sodium neglecting the contribution from the UV continuum short-ward of 13.6\unit{eV}. Including these photons erroneously ionizes hydrogen, but makes only a minor difference to the ionization rate of sodium. For photons with energies less than $13.6\unit{eV}$, we assumed the optical depth is small. The ionization fraction of hydrogen is also calculated in the neutral region, since, at apart from the highest densities, hydrogen provides the dominant contribution to the electron density. Since the ionization fraction can be very low in the densest parts of the wind we applied a floor to the electron fraction, $X_{\rm e} = n_{\rm e} / n_{\rm H} = 10^{-5}$, to represent the contribution from metals.

Away from the white dwarf, the dominant contribution to the hydrogen ionization rate comes from X-ray ionization, which was also included in the ionization calculation for sodium. The X-ray ionizing rate was computed using the fits given by \citet{Kozma1992}, which show that approximately 40 per cent of the X-ray energy deposited goes into ionizing the gas. The rate of X-ray energy deposited can be estimated as $F_X \sigma_X$, where $\sigma_X$ is the effective X-ray absorption cross-section and $F_X$ is the X-ray flux. This was estimated using the cross-section from \citet{Morrison1983}, averaged over a black-body with temperature $2 \times 10^{6}\unit{K}$. The observed luminosity is translated into a flux taking account of the fact that the X-ray emission comes from the shocked nova shells, which are external to the red giant wind. In this case, treating the emission as a point source would over estimate the flux near the binary. Instead the flux is approximated via $F_X = L_X / 4\pi \max(r,r_s)^2$, which gives the correct limits for $r \ll r_s$, where $r_s$ is the radius of the nova shell.  During quiescence we used $r_s = 10^{15}\unit{cm}$. During outburst $r_s$ evolves since the nova shells expand -- we used the mean radius of X-ray emitting gas, which is taken to be the gas with $T > 10^{6}\unit{K}$.  

During quiescence we calculated the ionization state in a steady-state approximation, using the same luminosities as for the Str\"omgren sphere calculations. For the line profile evolution after the nova the quiescent UV flux was used, since we considered the evolution of the lines after the end of the super-soft source phase. By this time accretion had resumed in RS Oph \citep{Worters2007}, so the quiescent UV flux is appropriate. For the soft X-rays, we used the observed luminosity from \citet{Bode2008}.  At the end of the super-soft phase, 100 days after the explosion, the luminosity was $5 \times 10^{35} \unit{erg}\unit{s}^{-1}$, decaying to the quiescent value by 1200 days after the explosion. During this time the ionization state was calculated time-dependently, assuming that both sodium and hydrogen were fully ionized at the end of the super-soft source phase and the temperatures were taken from the SPH simulation. We found the ionization fraction of sodium in the regions where hydrogen is neutral was in the range $10^{-3}$ to $10^{-4}$ during quiescence. For the novae, initially the high temperature in the nova shell prevents sodium from recombining. As the shells expand and cool, sodium in the equatorial ring is able to recombine. However, at lower inclinations the nova ejecta remains ionized. The effects of ionization can be seen in the line profiles presented in section \ref{Sec:Na}.

\section{Polar Inflow}
\label{Sec:Ha}
\begin{figure*}
\centering
\includegraphics[width=\textwidth]{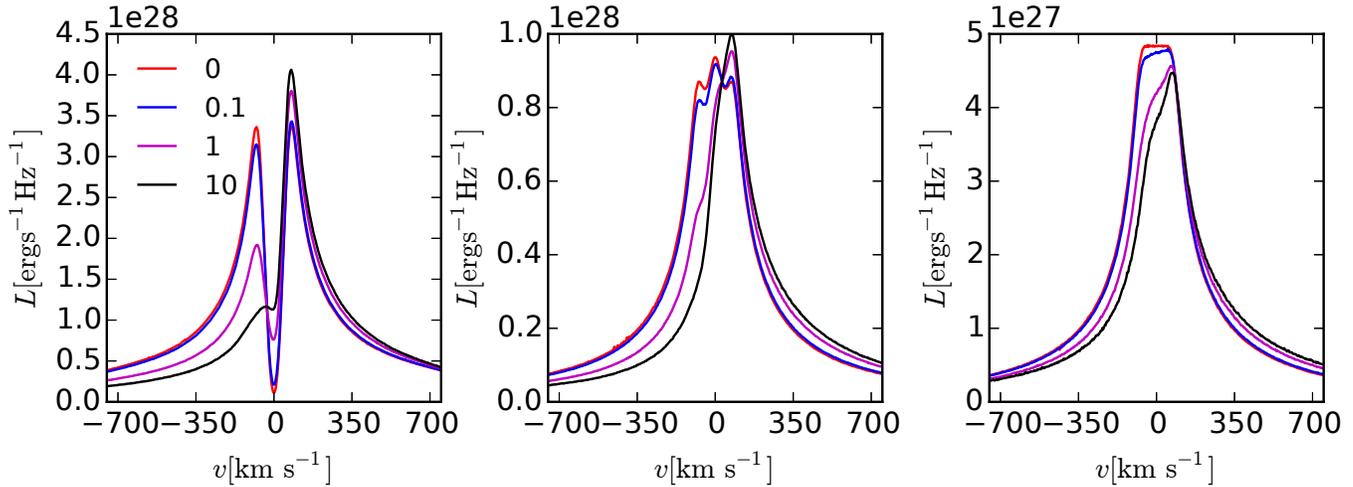}
\caption{H\,$\ualpha$ emission from the polar inflow model. The key denotes the scattering optical depth in the accretion disc and the observer is at an inclination of 45\degree{}. In each model the accretion rate is the same, but the opening angle of the polar inflow increases left to right across the panels with $\theta = 30\degree{}$, 60\degree{} and 90\degree{} respectively. \boldtxt{Note the different luminosity scales for the different opening angles.}}
\label{Ha_emiss}
\end{figure*}

The simulations show the presence of a polar inflow in the system, with gas falling back onto the binary from $10^{14}\unit{cm}$. The accretion rate in this component is approximately $5 \times 10^{-9}\unit{M}_{\sun}\unit{yr}^{-1}$ (Sec.~\ref{Sec:Quies}). Polar accretion flows are a ubiquitous feature of binary wind models, arising due to the passage of the white dwarf through the red giant wind (\boldtxt{this can be seen in the upper right panel of Fig.~\ref{QuiesStruct}}). 

The occurrence of such a flow is supported, at least at distances of $10^{14}\unit{cm}$, by the detection of red-shifted components in absorption lines of Na, Ca and K in  RS Oph \citep{Patat2011}. However, the structure of the inflow within $10^{12}\unit{cm}$ of the white dwarf is affected by the use of a sink in the simulation and is therefore not well known. It is also possible that the structure may be further complicated by the occurrence of winds from the accretion disc \boldtxt{(not included in our models)}, since, although RS Oph is below the Eddington luminosity, $L/L_\mathrm{Edd} \sim 10^{-2}$, line opacity maybe sufficient for RS Oph to drive a disc wind \citep{Drew2000}.

X-ray observations of RS Oph during quiescence tentatively support the presence of a polar accretion flow. \citet{Nelson2011} found that the hard X-ray emission can be explained by emission from shocked gas as it settles onto the white dwarf, and measured an accretion rate $2 \times 10^{-9}\unit{M}_{\sun}\unit{yr}^{-1}$. They found a maximum temperature of $6\unit{keV}$, which corresponds to a shock velocity $v \approx 1.7 \times 10^{3} \kms$. This velocity is close to the width of the H\,$\ualpha$ emission line wings in RS Oph \citep{Zamanov2005}, suggesting the H\,$\ualpha$ emission could arise in an inflow.

We estimated the H\,$\ualpha$ emission from the inflow using a simplified model in which the emission comes from gas in free-fall onto the white dwarf. We neglect the effects of angular momentum on the flow, except that we do not include emission from the parts of flow with $v > 1\,700\kms$, since it has been shocked to high temperatures and forms the X-ray emitting cooling flow. The inflow is taken to be spherically symmetric; however, we also considered an axis-symmetric conical structure, since the accretion disc must block part of the inflow.

The luminosity is assumed to be due to recombination, 
\begin{equation}
L(\nu) =  h \nu f \alpha(T) \int_V n(r)^2 \phi(\nu') \diff^3r.
\end{equation}
The line profile, $\phi(\nu)$, was taken to be the Doppler profile, the density set via mass conservation, $n = n_0 r_0^2 v_0/(r^2 v{\bf(r)})$, where $v^2(r) = -2GM(r^{-1}-r_0^{-1}) + v_0^2$. The frequency $\nu'$ is the rest-frame frequency and $\phi(\nu')$ is the Doppler emission profile. The subscript zero denotes quantities measured from the simulations at $r_0 = 2.5 \times 10^{12}\unit{cm}$, which  gives $n_0 = 5 \times 10^{-16}\unit{g}\unit{cm}^{-3}$ and $ v_0 = 80\kms$. For the temperature, we use $T = 30\,000\unit{K}$, which only affect the total luminosity. The emissivity from recombination has been approximated via $h \nu f \alpha(T) n(r)^2$, where $\alpha$ is the recombination coefficient and $f$ average number of H$\,\ualpha$ photons per recombination. The recombination coefficient used is that for Case B, in which recombination photons emitted with $hv > 13.6\unit{eV}$ are assumed to be immediately reabsorbed. In this case, $f \approx 0.5$ \citep{Storey1995}.

Self-absorption in the inflow has been neglected, however, electron scattering from a razor thin accretion disc was included. Emission from material on the far side of the disc is reduced by a factor $\exp(-\tau)$, where $\tau$ is the optical depth through the disc and scattering was treated in the single scattering approximation. The scattered luminosity, $L_s(\nu)$, is derived by considering the scattering of photons emitted at a point $\vec{r}'$ by the accretion disc at a point $\vec{r}$. For an emissivity, $\eta(\vec{r}',\nu') = h \nu f\alpha n^2(r') \phi(\nu')$, the flux arriving at a location $\vec{r}$ in the disc from a volume element $\diff^3r'$ is 
\begin{equation}
F = \frac{\eta(\vec{r}',\nu') \diff^3 r'}{4 \pi |\vec{r} - \vec{r'}|^2}.
\end{equation}
The probability that the flux is scattered through an angle $\theta_s$ to the the observer is $(1 - \exp(-\tau)) I(\theta_s)$, where ${I(\theta_s) \propto 1 + \cos^2(\theta_s)}$ is the angular part of the Thomson scattering cross-section and the factor $(1 - \exp(-\tau))$ takes into account the optical depth through the disc. The total luminosity can be written as

\begin{equation}
L_s(\nu) = (1 - \exp(-\tau))\int \frac{\eta(\vec{r}',\nu')}{4 \pi |\vec{r} - \vec{r'}|^2} I(\theta_s(\vec{r}', \vec{r})) \diff^3r' \diff^2 r,
\end{equation}
where the integral over $\diff^3r'$ is over the volume of the emitting region and the integral $\diff^2r$ is over the surface of the disc. For coherent scattering the relationship between the emitted and observed frequencies, $\nu' = \nu'(\nu, \vec{r}', \vec{r})$, was calculated by considering the Doppler shift between the frame of the emitting gas and the rest frame of the disc at $\vec{r}'$, followed by the shift between the frame of the disc and the observer. 

The resulting line profiles for a range of cone opening angles and optical depths are shown in Fig.~\ref{Ha_emiss}. In the conical models, the density in the flow has been adjusted so that each model has the same accretion rate and the observer is at an inclination of 45\degree{}. Models with negligible optical depth in the accretion disc, $\tau = 0$, produce symmetric line profiles making it impossible to tell whether the emission is due to an outflow or inflow. The presence of scattering breaks the symmetry, producing profiles with stronger red-shifted emission. 

Models with low conical opening angles produce double-peaked profiles and intermediate models can produce triple peaked emission, although scattering by the accretion disc results in a single peaked profile. We note that while the model with an opening angle $\theta=30\degree{}$ and $\tau = 1$ mimics a P-Cygni profile, the H\,$\ualpha$ profile in RS Oph cannot be produced by this mechanism alone. The reason for this is that in the centre of higher Balmer series lines, such as H\,$\ubeta$ and H\,$\ugamma$, show P-Cygni profiles in which the flux is below the level of the continuum, and absorption must therefore be present \citep{Bulla2013}.

The P-Cygni absorption in RS Oph cannot be associated with a polar inflow; since it is blue-shifted, it must arise in an outflow. The most likely explanation is that the absorption arises further out in the wind where the temperature is lower. This is compatible with observations of the 2006 nova in RS Oph, which show the narrow P-Cygni profile was still present 2 days after the nova \citep{Patat2011}. For the absorbing material to be outside the shock front at this time the distance to the absorbing material must be at least $5 \times 10^{13}\unit{cm}$.

Furthermore, unlike the H\,$\ualpha$ line wings, the velocity absorption component does not correlate with the motion of either the white dwarf or the red giant \citep{Brandi2009}. In contrast to this, the central velocity, as measured by the H\,$\ualpha$ line wings, does correlate with the motion of the white dwarf. This correlation would be expected for a polar inflow model since the line wings depend only on the accretion rate and white-dwarf mass, not the density and velocity away from the white dwarf. Therefore, we suggest the polar inflow as an explanation for the line wings. However, the central component may be dominated by an outflow.

While the association of H\,$\ualpha$ line wings with an inflow may be compelling, we note that accretion disc wind models may be able to produce similar line profiles. Calculations of lines formed in disc-winds from white dwarf stars have been limited to resonance lines tend to produce absorption profiles for inclinations $i \gtrsim 70\degree{}$ \citep{Proga2002}. However, disc wind models for T Tauri stars can produce similar H\,$\ualpha$ line profiles to those seen in RS Oph \citep{Kurosawa2006}. Polarization measurements may be able to distinguish between inflow and outflow models since, in the inflow models, the fraction of scattered light contributing to the blue wing is larger.

\section{Sodium Absorption Lines}
\subsection{Na \textsc{i}\,D lines in RS Oph}
\label{Sec:Na}
\begin{figure*}
\centering
\includegraphics[width=\textwidth]{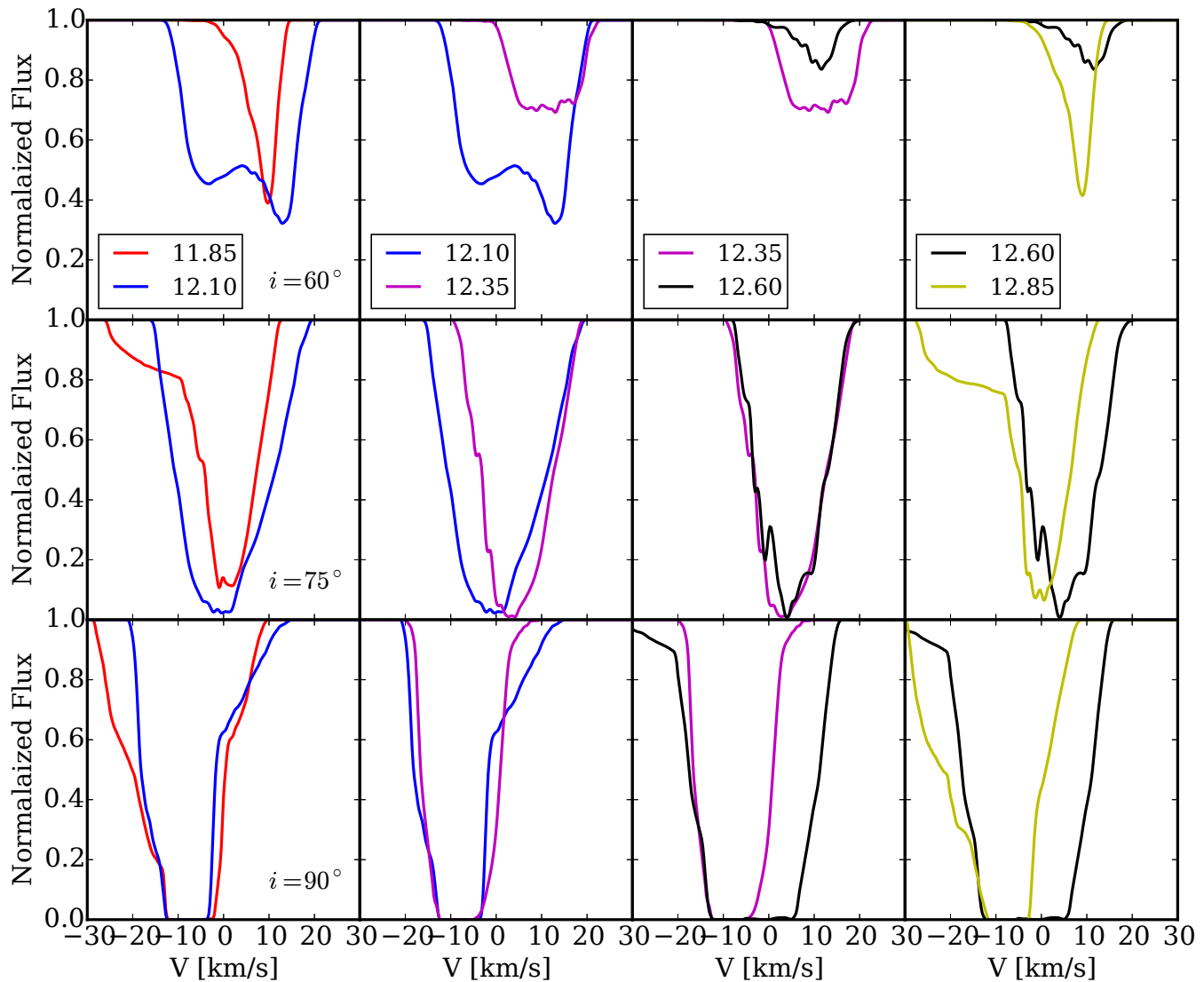} 
\caption{Model Na\,\textsc{i} D absorption lines produced during the quiescent mass-loss phase, at inclinations of $60\degree$ (top), $75\degree$ (middle) and $90\degree$ (bottom). The key denotes the number of orbits since mass transfer began {where the red giant is nearest to the observer at phase 0}. Since the blue-shifted components arise in the circumstellar outflow, they disappear at lower inclinations due to the lower column depth. The red-shifted component varies less strongly with inclination angle.}
\label{NaQuies}
\end{figure*}

\begin{figure*}
\centering
\includegraphics[width=\textwidth]{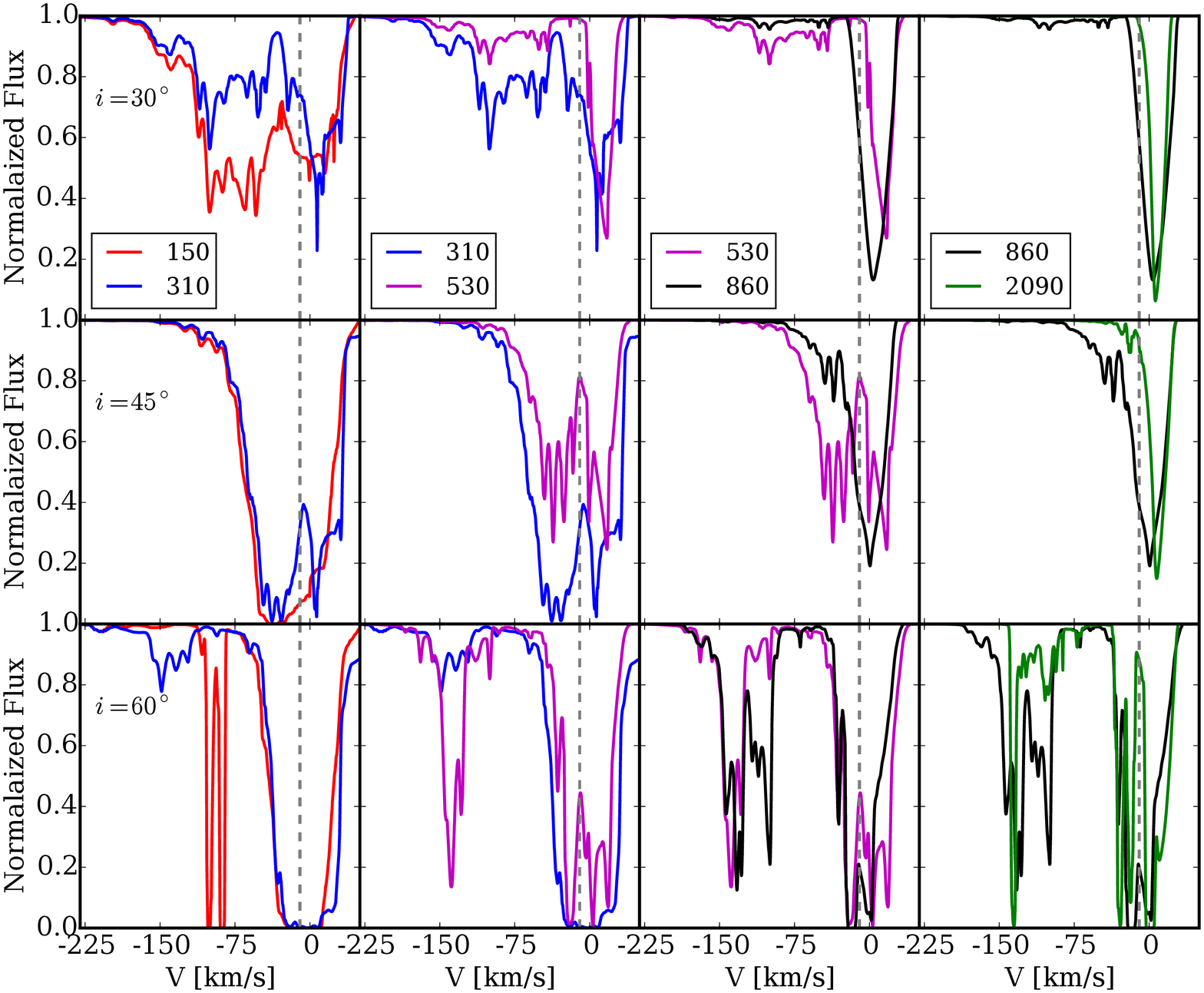}
\caption{Contribution of the nova ejecta to the Na\,\textsc{i} D lines for inclinations of 30\degree{} (top), 45\degree{} (middle) and 60\degree{} (bottom). The key denotes time in days since the start of the nova outburst. \boldtxt{The grey line marks the wind velocity $v \approx -12\kms$, column densities at velocities greater
than this will be affected by interactions with the RG wind}. For inclinations $i \lesssim 45\degree{}$, the high velocity components weaken as the nova shell evolves. At higher inclinations, for lines of sight intersecting the equatorial ring, a component at $150\kms$ appears once the ring has cooled below $10^3\unit{K}$. The components at $100\kms$ for $i=60\degree{}$ arise in the gas evaporated from the accretion disc. The evaporative component weakens continually after the nova, and may explain the additional material seen after the 2006 outburst of RS Oph.}
\label{NaEvo}
\end{figure*}

Sodium absorption lines have been suggested as a useful diagnostic of the circumstellar environment in both supernovae and their progenitor systems \citep{Patat2007,Chugai2008,Patat2011}. The Na\,\textsc{i} D lines, together with other strong metal lines such as the Ca\,\textsc{ii} H \& K lines, probe the density and ionization conditions in the circumstellar environment. We modelled the line profiles expected from the simulations, with the aim of understanding where the observed features in RS Oph form in the CSM and whether the same circumstellar components can be responsible for metal line variations observed in some SN Ia.

We have modelled Na\,\textsc{i} D absorption line profiles directly from the simulations, using the results from the ionization calculation described above. All neutral sodium atoms are assumed to be in the ground state and contribute to the absorption lines. We present line profiles from calculations in which the mass-loss rate has been scaled to $5 \times 10^{-7}\unit{M}_{\sun}\unit{yr}^{-1}$, since this gives better agreement with observations both for the ionization structure and the line profiles. We did not include a detailed treatment for the H\,\textsc{ii} region in the Na\,\textsc{i} D line ionization calculation; instead we neglected the contribution from all material within $3\times 10^{13}\unit{cm}$. The results are not particularly sensitive to this choice of radius.

We assume that the continuum flux around the Na\,\textsc{i} D lines is dominated by the red giant. The presence of an emission component in RS Oph limits the extent to which this is true \citep{Patat2011}. The emission component has a width $v \sim 100\kms$, higher than the width of the absorption lines, and is likely associated with emission from the ionized region. The line profiles were  derived by calculating the optical depth to the photosphere along lines of sight through the CSM. The photosphere was subdivided into lines of sight to ensure that the sub-structure along the line of sight is resolved. Finally, the flux from each line of sight was combined and weighted by area.

Fig.~\ref{NaQuies} shows the resulting synthetic line profiles. For inclinations $i \gtrsim 70\degree{}$, the line profile shows both red- and blue-shifted components. The red-shifted components arise in the wind that is falling back onto the white dwarf and accretion disc, and is therefore strongest when the white dwarf is closest to the observer and the line of sight intersects the most material. The blue-shifted components arise further out in the CSM, beyond $10^{14}\unit{cm}$, which has a terminal velocity of 10 to 20\kms. At higher inclinations this component is relatively stable with phase \boldtxt{because the distant material dominates the line profile and contributes to all lines of sight independent of phase. At low inclinations  the contribution from material at large distances is smaller, due to the lower density far above the mid-plane. Therefore, the high inclination line profiles are more sensitive to the material close to the binary, which varies with phase.

The inclination at which the behaviour of the blue-shifted component changes from stable to phase-dependent
depends on how concentrated the outflow is. For our cooling model (Fig.~\ref{NaQuies}) this occurs at $i \sim 70\degree{}$, but for models without cooling the line profile is stable for a larger range of inclinations, with the transition at $30\degree{}$ to $45\degree{}$. In RS Oph the blue-shifted lines are relatively stable, which suggests that the outflow in RS Oph are less strongly concentrated than in our cooling model, since the inclination is $i\approx 50\degree{}$.}

The absorption lines formed in the nova shells are shown in Fig.~\ref{NaEvo}. These lines were modelled from the end of the super-soft source phase and assume that the central system had reached a quiescent steady-state once more. The evolution of the line profiles is governed by the competing effects of recombination, which strengthen the lines, and expansion which weakens them.  \boldtxt{Since the nova shell expands rapidly in the polar directions this component is not dense enough to form strong sodium lines (at $3$ to $4\times10^3\kms$). Instead, the absorption line profiles at $i \lesssim 60$ are dominated by the `evaporative wind', which has $v \lesssim 100\kms$. The large width of the line for $i = 30\degree{}$ is a result of the evaporative wind  lying along the line of sight to the observer, while at higher inclinations
lines of sight cut across the evaporative wind.}

For inclinations of $i \gtrsim 60\degree{}$, the interaction of the nova shell with the wind is strong enough to produce an additional component to the line profile, at higher velocity than the material from the accretion disc. Initially the continued interaction of the nova shell with the wind keeps the shell hot, with collisional ionization preventing recombination in the sodium. As the shell eventually cools and recombines strong absorption lines appear.  While the components from the evaporative wind weaken over time, they can still be seen 2000 days after the nova.

The different absorption components in our models can be used to determine the origin of the absorption components observed in RS Oph. The gas that falls back onto the binary from above and below the plane naturally explain the otherwise unexplained red-shifted absorption components seen in RS Oph. Additionally we find the terminal velocity in the outflow is 10 to 20$\kms$, which is considerably lower than the fastest observed component, which has a velocity $v = - 37 \kms$. While additional sources of acceleration could be invoked to explain the difference, the most natural explanation comes from the novae, which show the evaporation of gas from the accretion disc after the nova has passed at velocities of $v \sim - 50\kms$.

Observations before and after the 2006 outburst support the suggestion of differing origin for the 10 to $20\kms$ and $37\kms$ components. Since after the outburst the lower velocity component was weaker, while the higher velocity component was stronger \citep{Patat2011}. The strengthening of the high-velocity component is expected if it is associated with the evaporative wind, as it expands and the lines weaken continually as the nova shells evolve. Similarly, lines arising in the wind should be weaker after the nova as the nova shells sweep up the wind, which slowly replenishes once accretion resumes. Observations of H\,$\ualpha$ and H\,$\ubeta$ absorptions prior to the 2006 nova support this, since they had the same velocity as the slow Na\,\textsc{i}\,D component, $v \sim - 10\kms$, and were observed to continuously strengthen during the period prior to the 2006 outburst \citep{Brandi2009}. Therefore, we suggest a combined quiescent mass-loss and nova model provides a natural explanation for the lines observed in RS Oph.

\subsection{RS Oph and Type Ia supernovae}

\begin{figure*}
\centering
\includegraphics[width=\textwidth]{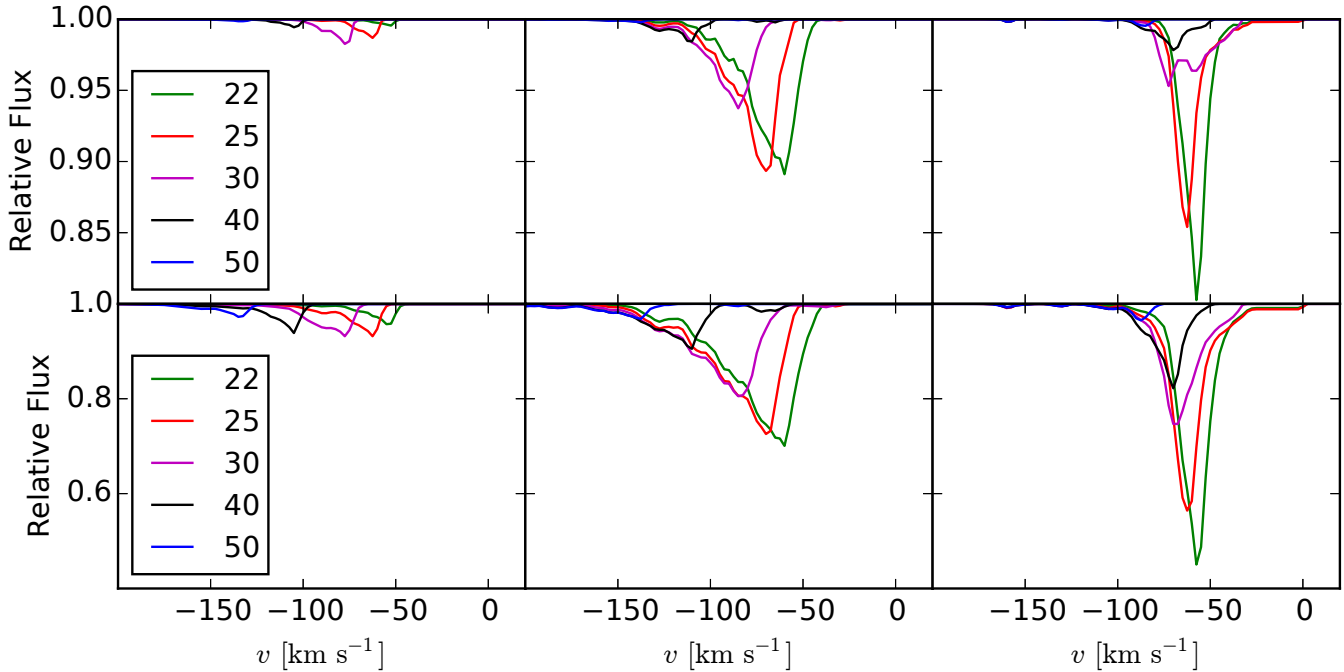}
\caption{Model Na\textsc{i}\,D lines as seen against a supernova photosphere. The key denotes days since explosion. \emph{Top panels:} Line profiles for the high mass-loss rate model that best matches RS Oph, at inclinations of $30\degree$, $60\degree$ and $90\degree$ (left to right). \emph{Bottom panels:} The same, but with the density increased by a factor of 30. The evolution is due to the combined effects of recombination, which is assumed to begin 20 days after the explosion, and the supernova sweeping up the CSM, continually removing the lower velocity material.}
\label{Na_SN}
\end{figure*}

In order to compare our simulations of RS Oph with observations of Type Ia supernovae, we calculated the line profiles as would be seen against a supernova photosphere. For the ionization conditions in the supernova,  we follow the hypothesis suggested by \citet{Patat2007} for explaining the observations of SN 2006X -- the supernova initially ionizes the circumstellar medium, which then begins to recombine 20 days after the explosion. The supernova is taken to be an expanding sphere with a velocity of $20,000\kms$, which sweeps up the circumstellar medium, and the for the photosphere we use a simple approximation due to \cite{Chevalier1981} is used
\begin{align}
R_\mathrm{p} = \begin{cases} 2.61 \times 10^{14} t_\mathrm{d}^{2/3} \unit{cm} & \qquad t_\mathrm{d} < 40.7, \\ 
							8.86 \times 10^{13}  \left[1 - \left(\frac{t_\mathrm{d}}{108}\right)^2\right] t_\mathrm{d} \unit{cm} & \qquad t_\mathrm{d} > 40.7, \end{cases} 
\end{align}
where $t_d$ is the time in days. This expression overestimates the photospheric velocity $v_\mathrm{p} = R_\mathrm{p} / t$ by approximately 25 per cent when compared with photospheric velocities from abundance tomography \citep{Stehle2005, Mazzali2008}. The supernova photosphere begins to recede after approximately 60 days, and the supernova starts to become optically thin to continuum radiation and a model for the emissivity should be used instead of an approximate photosphere. However this does not occur until after the line profiles have finished evolving.

We assume the explosion occurs after  the end of the simulations, once three recurrence times occur, which corresponds to the lowest density of, and largest the distance to the inner-most nova shell. While there is no reason for the supernova to occur at this specific time, there is additional uncertainty in recurrence time and mass-loss rate prior to the supernova, which also affect the distance to, and density of the shells. The resulting line profiles are shown in Fig. \ref{Na_SN}. For our model that uses the density directly from the simulations of RS Oph, the density in the nova shell is not high enough that for the sodium to significantly recombine. However, since the simulations may underestimate the density in the shell, we also include a model in which the recombination time-scale has been reduced by a factor of 30, which corresponds to an equivalent increase in density, but the same total mass. 

For supernovae that occur in systems that have low inclinations, $i \lesssim 45\degree$, circumstellar absorption lines are likely to be undetectable. For higher inclinations, in which lines of sight intersect the high density gas in equatorial ring the line profiles show a characteristic evolution. The evolution is due to the competing effects of recombination and the sweep up of circumstellar gas by the supernova. Since in general the most dense material is at the lowest velocities, it both recombines first and is swept up first. In addition to recombination -- sweep-up evolution, the covering factor can also play a role in the line profiles. In these models, the dense equatorial ring only covers the whole photosphere when the system is close to edge on, other wise the covering factors are smaller and the absorption lines will never be saturated, regardless of the shell mass.

In addition to the distance to the shells, the time that it takes for a given component to disappear is dependent on the thickness of the shell. This effect can be seen in Fig. \ref{Na_SN}, since the effective thickness at $i = 60\degree$ is larger than that at $i = 90\degree$. For SN 2006X, the lowest velocity component disappeared between days $+14$ and $+61$ \citep{Patat2007}. Taking a velocity of $20,000\kms$, this gives a thickness of $8 \times 10^{15}\unit{cm}$, although since the component was not observed during its disappearance, the shell could be considerably narrower. For comparison, the simulated nova shells have a thickness of $2 \times 10^{15}\unit{cm}$. While this may be an over-estimate of the thickness, it is well within the observational limit.

The model line profiles reproduce a number of other qualitative features of SN 2006X - i.e. the lower velocity components are stronger and recombine first, and are closer to supernova and therefore also disappear first. Only the gas trapped in the dense shells produces significant absorption components, which requires a dense circumstellar environment for the nova shells to sweep up, hence the lack of absorption lines at an inclination of $0\degree$. The comparable strength of the $i = 45\degree$ and $ i= 90 \degree$ line profiles in spite of the much higher density in the equatorial plane can be explained by the relatively short distance to the nova shell, $r \sim 3 \times 10^{15}\unit{cm}$, which means that by 20 days the supernova is already starting to sweep up the shell.

For nova shells at larger distances, competing factors will affect the line profile. For example, at the same mass-loss rate longer recurrence periods will lead to larger distances and more massive shells. If the recurrence period was 10 times longer, the shells distance to the shells would be $10^{16}$ to $10^{17}\unit{cm}$, which is closer to the distance estimated for SN 2006X \citep{Patat2007,Chugai2008}. At fixed mass, the expansion of the shells reduces the density by a factor of 100 to 1000, depending on whether the shells continue to expand at constant velocity, or whether a significant velocity gradient develops in the shell. Taking into account the additional mass loss associated with the longer recurrence time, the density would be expected to be 10 to 100 times lower, which would produce lines that are at least a factor of 10 weaker.

However, \citet{Chugai2008} suggested that the supernova would continue to ionize neutral sodium even after maximum light, despite the low UV flux since there is a two-photon process that uses optical photons. Conversely, the ionization of hydrogen is much less effective due to UV line blocking, with the hydrogen ionizing flux dominated by $\gamma$-ray emission and X-ray emission from circumstellar interaction. In this case a density $n_{\rm H} > 10^{7} \unit{cm}^{-3}$ at several $10^{16}\unit{cm}$ may be required, considerably above the peak density found in the simulations $\sim 10^{5}\unit{cm}^{-3}$. 

Under these circumstances it is possible to estimate the dependence of the line strength on the mass-loss rate and recurrence time. Since the ionization fraction for hydrogen $X_H \ll 1$, it can be estimated as $X_{\rm H} = \sqrt{I_{\rm H} / \alpha_{\rm H} n_{\rm H}}$, where $ I_{\rm H} = \langle F_{\rm H} \sigma_{\rm H}\rangle \propto r^{-2}$ is the effective ionizing flux deposited in the shell and $\alpha_H$ is the recombination rate. Similarly, since sodium is nearly completely ionized, the fraction of sodium atoms that are neutral, $Y_{\rm Na}$, can be estimated as $Y_{\rm Na} = 1 - X_{\rm Na} = I_{\rm Na} / \alpha_{\rm Na} n_e$, where $n_e$ is the number density of electrons $n_e \approx X_{\rm H} n_{\rm H}$. The difference in the square root between the two expressions arises since both the number of electrons and ionized hydrogen atoms are proportional to $X_{\rm H}$, which both appear in the recombination rate, but only the number of ionized sodium atoms depends on $X_{\rm Na}$.  For the ionization of sodium via the two photon process, $I_{\rm Na} \propto r^{-4}$, and $Y_{\rm Na} \propto n_{\rm H}^{1/2} r^3$. This shows that, if the two-photon process is effective at ionizing sodium, then a longer recurrence time will lead to stronger absorption lines. However, this would require more mass in the circumstellar environment than the case in which two-photon ionization is ineffective. The range of parameter space favoured by our models for explaining the absorption lines in SN 2006X is therefore a mass-loss rate of order $10^{-6}$ to $10^{-5}\,M_\odot\unit{yr}^{-1}$ and a recurrence time of a few tens to a few hundreds of years.

\section{Discussion}
\label{Sec:Discussion}
We have simulated the structure of the CSM in RS Oph, during quiescence and after multiple novae explosions.  We find evidence for accretion onto the white dwarf from the polar directions, which may explain the broad wings of the H\,$\ualpha$ emission line. Using the structure from simulations, we have modelled the ionization conditions in the CSM, and by calculating synthetic line profiles have deduced the origin of observed features in the Na\,\textsc{i} D lines. Additionally we have modelled the line profiles that might be observed if a Type Ia supernova occurs in a system that is similar to RS Oph.

The simulations of the quiescent mass-transfer phase produce a highly equatorial wind, tightly focused towards the binary orbital plane. The outflows are  more focused than the simulations conducted by \citet{Walder2008}. The difference likely arises because they assumed the inverse of the mass ratio, i.e., a low mass white dwarf and a heavier giant, and also due to the combination of both the lower wind velocity we assume, chosen to mimic Roche lobe overflow of the stellar atmosphere; and a different equation of state, \citet{Walder2008} used a near-isothermal equation of state, chosen to approximate the heating and cooling balance due to photoionization. Our calculations directly included radiative cooling, but not heating from photoionization, resulting in lower temperatures and a more focused outflow. 

The accretion disc and the equatorial wind give rise to a  highly bipolar nova structure, including a dense equatorial ring, which differs from the elliptical structure produced in the calculations by \citet{Walder2008}. While neglecting heating undoubtedly produces a more focused wind than in RS Oph, \textit{HST} imaging of the 2006 outburst shows tentative evidence for a bipolar structure and equatorial ring \citep{Bode2007}. The circumstellar structure in RS Oph is most likely somewhere between our models  and that of \citet{Walder2008}. The differences in the equation of state that \citet{Walder2008} used an our models have been assessed by running models with suppresses cooling. Since these models produce winds with increased vertical extent, but still produced bipolar novae, the more important differences are the assumptions of the observed mass ratio and the lower wind velocity used in our calculations.  Since a low wind velocity is required to reproduce the bipolar structure in the novae, accretion in RS Oph must occur from low velocity material that is filling the Roche lobe, rather than through standard Bondi-Hoyle-Littleton wind accretion \citep{Edgar2004}.

The models capture many of the features of Na\,\textsc{i} D absorption line profiles observed in RS Oph, providing an explanation for the red-shifted components to the Na\,\textsc{i} D and Ca\,\textsc{ii} H \& K lines. We find blue shifted components at velocities of $-10$ to $-20\kms$ that form in the wind, along with the evaporative wind at velocities of approximately $-50\kms$, which together may explain the absorption components observed in RS Oph. The respective weakening and strengthening of the $-10$ and $-37\kms$ after the 2006 outburst support the differing origins of these components. If our hypothesis is correct then low velocity components must strengthen, while the high velocity components should have weakened again. Furthermore, if the high velocity components arise  in the evaporative wind it should have similar composition to the quiescent mass loss, rather than the nova ejecta which are processed in the outburst.

As well as the components at  $v \gtrsim -50\kms$, the nova line profiles show the presence of strong absorptions components at higher velocities. These components form in the shell swept up by the novae and are visible for inclinations $i \gtrsim 60\degree{}$ and remain visible at the time subsequent novae occur. The velocity of the shell depends on viewing angle, $v \approx -150\kms$ for $i = 60\degree{}$ reducing to $v \approx -50 \kms$ for $i = 90\degree{}$. While this material is not seen in absorption in RS Oph, it provides the most promising prospects for the origin of absorption lines observed in some SNe Ia.

\begin{figure}
\centering
\includegraphics[width=0.45\textwidth]{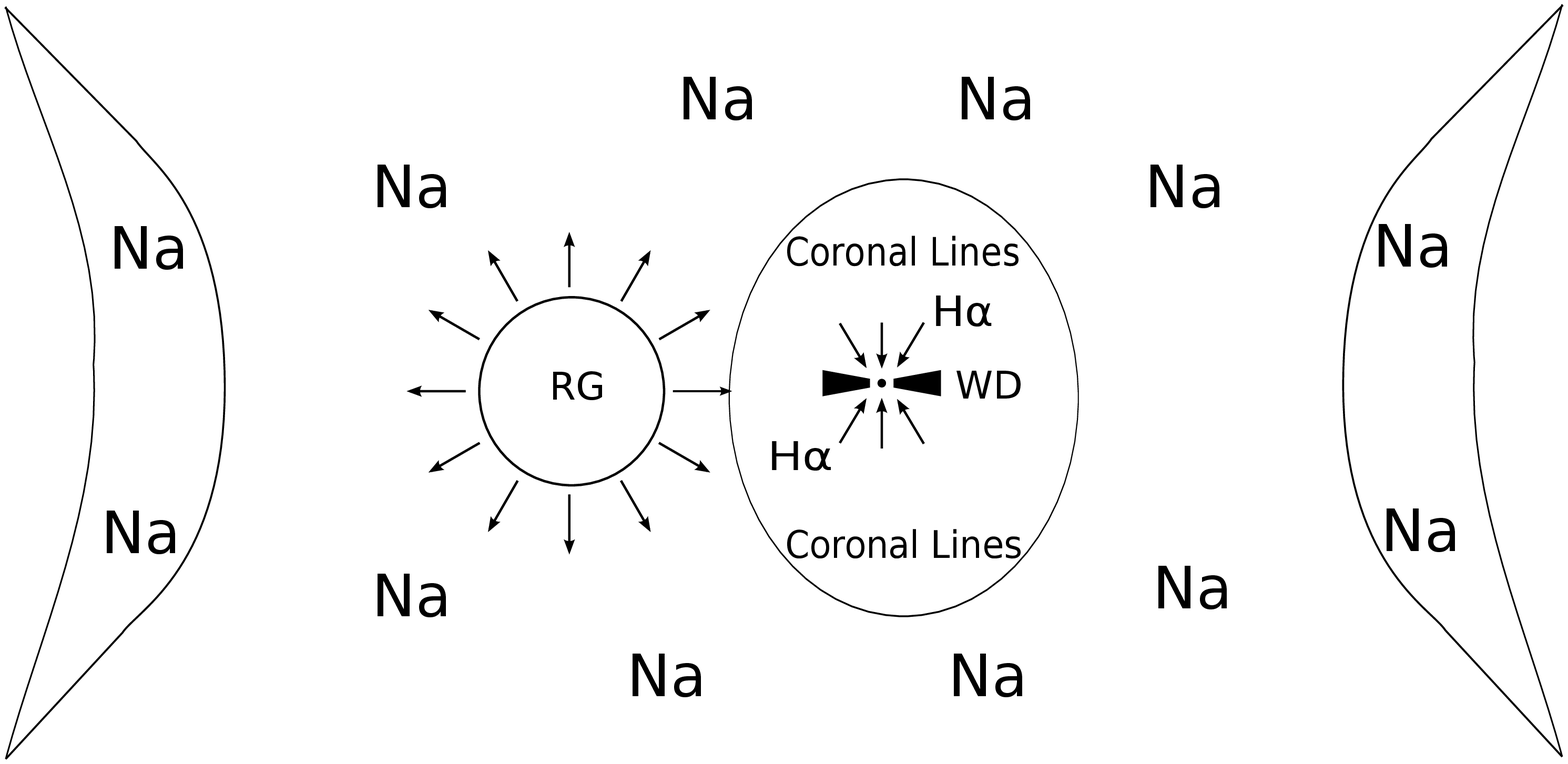}
\caption{Schematic for line formation during the quiescent mass-transfer phase in RS Oph: H\,$\ualpha$ emission occurs close to the WD, in material falling onto the poles of the WD. The P-Cygni type absorption is formed in the photoionized wind, close to the binary, along with the coronal cF-absorption \citep{Brandi2009}. The Na\,\textsc{i} D and Ca\,\textsc{ii} H \& K lines form in the neutral part of the wind, with the stable blue-shifted components forming far out, and the red-shifted components forming in the gas falling back onto the binary. The nova shells also contribute to the absorption lines for high inclinations, and it is this material that may be observed in SNe Ia.}
\label{Cartoon}
\end{figure}

In the event of a supernova occurring in an RS Oph-like system, the nova shell components at $-50$ to $-150\kms$ should be visible right up to the explosion. We have modelled resulting line profile evolution assuming the ionization undergoes an ionization -- recombination cycle, similar to SN 2006X \citep{Patat2007}. The line profiles are characterised by the lowest velocity material recombining first, which is swept up by the supernova in the 20 days after maximum light. This behaviour is similar to the lowest velocity component in SN 2006X, however, a higher mass-loss rate of order $10^{-6}$ to $10^{-5}\,M_\odot\unit{yr}^{-1}$ may be necessary to be consistent with photoionization modelling \citep{Chugai2008,Simon2009}.

Such high mass-loss rates appear to be in contention with radio and X-ray observations, both of Type Ia supernovae in general, but also of SN 2006X itself in which there was no detection \citep{Stockdale2006,Panagia2006}. The upper limits for SN 2006X appear to limit a mass-loss rate to less than a few $10^{-8}\,M_\odot\unit{yr}^{-1}$, and in the case of a few other supernovae such as SN 2011fe or SN  2014J the limits appear to be even more stringent \citep{Chomiuk2012,Perez-Torres2014}. While it is clear that the circumstellar medium in some of these objects must be very diffuse, this is incompatible with the Na\,\textsc{i}\,D line absorption in SN 2006X, which requires a significant amount of mass within $10^{17}\unit{cm}$. Reconciling these conflicting results is clearly important for understanding the nature of SNe Ia more generally, since approximately 20 per cent of them show evidence for out-flowing material \citep{Maguire2013}.

This work highlights one possible explanation for the discrepancy between radio limit and absorption lines, which is related to the highly aspherical geometry associated with circumstellar outflows and novae. Most of the mass loss is confined to the equatorial plane, resulting in dense ring like structures after nova explosions. Both of these features will undoubtedly produce different radio and X-ray emission to a spherical wind model, perhaps with the emission confined to a narrow time window when the supernova is interacting with the circumstellar shells. This is especially true early on, while the radio emission remains optically thick and therefore can possibly hide a large fraction of the mass \citep{Chevalier1998,Chomiuk2012}.  A thorough investigation of these differences deserves merit, but is deferred to a future study.

Another possible way to reconcile the difference between the large the mass required and the low mass-loss rate limits is that the absorption lines form as a consequence of the build of mass loss over a long time. One particular way this could occur is through the pile up of many nova shells. Prior to the nova phase, the progenitor should be surrounded by a dense wind or interstellar medium that continually slows down the first nova shell. Since subsequent novae are only slowed down by the inter-outburst mass loss, the shells will eventually catch up with each other, forming a massive shell which remains even at large distances from the progenitor \citep{Badenes2007,Dimitriadis2014}. In this way it may be possible to achieve sufficient densities to explain the narrow absorption lines without requiring large mass-loss rates. Since termination the shell is formed by the novae sweeping up a large mass, the shells should have velocities of a few 10s of \kms and the large distance to these termination shells ($\gtrsim 0.1\unit{pc}$, for a $10^4\unit{yr}$ nova phase) suggests that time-dependence should be weak. 

The role of the interstellar medium in the formation of these shells may explain the fact that no narrow absorption lines have been seen in SNe Ia hosted by elliptical galaxies \citep{Maguire2013}, without invoking vastly different progenitors between galaxy types. For SNe Ia produced by the WD + MS channel, which are expected to make up the majority of SNe Ia from the single degenerate progenitors, the mass-loss rates are considerably lower than symbiotic binaries \citep{vandenHeuvel1992,Li1997,Han2004}. Therefore, the formation of termination shells may require a relatively dense interstellar medium to slow down the nova shells, which is not present in elliptical galaxies. Conversely differing progenitor channels in different galaxy types is already suggested by variations in the intrinsic properties of SNe Ia in different galaxy types \citep{Hamuy1995,Cappellaro1997,Sullivan2006}, a hypothesis that would be strengthened if narrow absorption lines are found in SNe Ia hosted by elliptical galaxies, with the properties of these SNe matching more closely those seen in spiral galaxies. 

In Fig.~\ref{Cartoon} we show a schematic of where the observed lines form in RS Oph. For SNe Ia, only the lines forming in the nova shells are relevant. For lines observed in RS Oph, the Na\,\textsc{i} D lines form furthest out, producing stable profiles that show some variation with phase. Within the gas producing the Na\,\textsc{i} D lines, the red-shifted components form closest in, in material falling back from above the plane of the binary. The wings of H\,$\ualpha$ line form close to white dwarf, as shown by their correlation with the motion of the white dwarf \citep{Brandi2009}, either in the polar inflow or a disc-wind. However, the P-Cygni absorption forms further out, where the gas is out-flowing. This is supported by observations of novae, in which the P-Cygni line was still present two days after the outburst, suggesting the material is at least $10^{14}\unit{cm}$ from the white dwarf \citep{Patat2011}. This is further supported by the lack of a correlation with the velocity of either star \citep{Brandi2009}. Also present in RS Oph are many absorption lines usually found in the spectra of supergiant stars, so called cF-absorption lines. Since the cF-absorptions also do not correlate with the motion of either component, the most likely origin for these lines is the ionized region close to the binary \citep{Dobrzycka1994,Brandi2009}. 

\section{Conclusions}

Understanding the circumstellar environment around supernovae provides a unique way of linking supernovae to their progenitors. This is especially true for Type Ia supernovae, where there is difficulty in obtaining direct detection of progenitors in pre-explosion images and where a determination of the circumstellar environment may discriminate between competing progenitor scenarios. Since the presence of a CSM has been detected in recurrent novae \citep{Williams2008,Patat2011}, but in general is not expected for double-degenerate systems, observations may be able to indirectly infer the progenitors from the pre-explosion CSM. However, some recent models predict a CSM in certain types of double-degenerate systems \citep{Shen2013}.

Using Smoothed Particle Hydrodynamics, we have modelled the quiescent mass-transfer phase in RS Oph, and the environment formed by the subsequent interaction of multiple nova ejections with the highly aspherical circumbinary medium. We find evidence for peculiar modes of accretion, with the simulations producing an eccentric accretion disc along with accretion from the polar directions. The polar accretion flow may be responsible for the observed hard X-ray emission and broad H\,$\ualpha$ wings associated with the white dwarf in RS Oph. We find that accretion disc survives the outburst, but is heated by shocks from the interaction with the nova ejecta. The excess thermal energy drives an evaporative flow from the accretion disc after the nova has passed.

The interaction between the nova and the quiescent mass-loss results in the formation of a bipolar structure with a dense equatorial ring. The most natural explanation for the origin of the Na\,\textsc{i} D and Ca\,\textsc{ii} H \& K absorption lines seen in RS Oph comes from a combination of the wind and the evaporative flow, since, except close to the binary plane, the novae remain at high velocities and low densities. At higher inclinations, the equatorial ring produces absorption lines with velocities $-50$ to $-150\kms$. Although this material is not seen in RS Oph, it may be the origin of the lower velocity time-variable components observed in SN 2006X.

For an RS Oph-like CSM to explain the excess blue-shifted absorption seen in 20 per cent of SNe Ia, a long recurrence time and high mass-loss rate may be needed, which appears to be in contention with radio estimates. Instead, these absorptions may be attributed to the pile up of nova shells as they interact with the interstellar medium at large distances from the progenitor. Such a model should produce absorption lines with velocities of $-10$ to $-30\kms$, at distances of order $0.1\unit{pc}$. In this way, the lack of narrow absorption lines seen in SNe Ia hosted by elliptical galaxies may be interpreted as a result of the interstellar medium being too diffuse to slow down the nova shells sufficiently. In this way, even with a relatively low mass-loss rate at the time of explosion, a recurrent nova model may be able to explain the blue-shifted absorption lines seen in SNe Ia.

\section*{Acknowledgements}
We thank F. Patat, B. Espey and J. Mikolajewska for many useful discussions relating to this work. We thank the reviewer for their comments which helped improve the clarity of the paper. SM acknowledges the receipt of research funding from the National Research Foundation (NRF) of South Africa. Figs.~\ref{QuiesStruct},  \ref{AdiabatStruct} and \ref{Novae} were made using \textsc{pynbody} \citep{pynbody}, while \ref{Ioniz} were made using SPLASH, a publicly available SPH visualization tool \citep{Price2007}.

\footnotesize{
  \bibliographystyle{mn2e_long}
  \bibliography{RSOph}

\begin{thebibliography}{79}
\expandafter\ifx\csname natexlab\endcsname\relax\def\natexlab#1{#1}\fi

\bibitem[{{Anupama}(2008)}]{Anupama2008}
{Anupama} G.~C., 2008, in Astronomical Society of the Pacific Conference
  Series, Vol. 401, RS Ophiuchi (2006) and the Recurrent Nova Phenomenon,
  {Evans} A., {Bode} M.~F., {O'Brien} T.~J., {Darnley} M.~J., eds., p. 251

\bibitem[{{Anupama} \& {Miko{\l}ajewska}(1999)}]{Anupama1999}
{Anupama} G.~C., {Miko{\l}ajewska} J., 1999, \aap, 344, 177

\bibitem[{{Arnaud} \& {Rothenflug}(1985)}]{Arnaud1985}
{Arnaud} M., {Rothenflug} R., 1985, \aaps, 60, 425

\bibitem[{{Badenes} {et~al}\mbox{.}(2007){Badenes}, {Hughes}, {Bravo}, \&
  {Langer}}]{Badenes2007}
{Badenes} C., {Hughes} J.~P., {Bravo} E., {Langer} N., 2007, \apj, 662, 472

\bibitem[{{Blondin} {et~al}\mbox{.}(2009){Blondin}, {Prieto}, {Patat},
  {Challis}, {Hicken}, {Kirshner}, {Matheson}, \& {Modjaz}}]{Blondin2009}
{Blondin} S., {Prieto} J.~L., {Patat} F., {Challis} P., {Hicken} M., {Kirshner}
  R.~P., {Matheson} T., {Modjaz} M., 2009, \apj, 693, 207

\bibitem[{{Bode} {et~al}\mbox{.}(2007){Bode}, {Harman}, {O'Brien}, {Bond},
  {Starrfield}, {Darnley}, {Evans}, \& {Eyres}}]{Bode2007}
{Bode} M.~F., {Harman} D.~J., {O'Brien} T.~J., {Bond} H.~E., {Starrfield} S.,
  {Darnley} M.~J., {Evans} A., {Eyres} S.~P.~S., 2007, \apjl, 665, L63

\bibitem[{{Bode} {et~al}\mbox{.}(2008){Bode}, {Osborne}, {Page}, {Beardmore},
  {O'Brien}, {Ness}, {Starrfield}, {Skinner}, {Darnley}, {Drake}, {Evans},
  {Eyres}, {Krautter}, \& {Schwarz}}]{Bode2008}
{Bode} M.~F. {et~al.}, 2008, in Astronomical Society of the Pacific Conference
  Series, Vol. 401, RS Ophiuchi (2006) and the Recurrent Nova Phenomenon,
  {Evans} A., {Bode} M.~F., {O'Brien} T.~J., {Darnley} M.~J., eds., p. 269

\bibitem[{{Brandi} {et~al}\mbox{.}(2009){Brandi}, {Quiroga}, {Miko{\l}ajewska},
  {Ferrer}, \& {Garc{\'{\i}}a}}]{Brandi2009}
{Brandi} E., {Quiroga} C., {Miko{\l}ajewska} J., {Ferrer} O.~E.,
  {Garc{\'{\i}}a} L.~G., 2009, \aap, 497, 815

\bibitem[{{Bulla}(2013)}]{Bulla2013}
{Bulla} M., 2013, Master's thesis, University of Milan

\bibitem[{{Cappellaro} {et~al}\mbox{.}(1997){Cappellaro}, {Turatto},
  {Tsvetkov}, {Bartunov}, {Pollas}, {Evans}, \& {Hamuy}}]{Cappellaro1997}
{Cappellaro} E., {Turatto} M., {Tsvetkov} D.~Y., {Bartunov} O.~S., {Pollas} C.,
  {Evans} R., {Hamuy} M., 1997, \aap, 322, 431

\bibitem[{{Chevalier}(1981)}]{Chevalier1981}
{Chevalier} R.~A., 1981, \apj, 246, 267

\bibitem[{{Chevalier}(1998)}]{Chevalier1998}
{Chevalier} R.~A., 1998, \apj, 499, 810

\bibitem[{{Chomiuk} {et~al}\mbox{.}(2012){Chomiuk}, {Soderberg}, {Moe},
  {Chevalier}, {Rupen}, {Badenes}, {Margutti}, {Fransson}, {Fong}, \&
  {Dittmann}}]{Chomiuk2012}
{Chomiuk} L. {et~al.}, 2012, \apj, 750, 164

\bibitem[{{Chugai}(2008)}]{Chugai2008}
{Chugai} N.~N., 2008, Astronomy Letters, 34, 389

\bibitem[{{Das}, {Banerjee} \& {Ashok}(2006){Das}, {Banerjee}, \&
  {Ashok}}]{Das2006}
{Das} R., {Banerjee} D.~P.~K., {Ashok} N.~M., 2006, \apjl, 653, L141

\bibitem[{{Dimitriadis}, {Chiotellis} \& {Vink}(2014){Dimitriadis},
  {Chiotellis}, \& {Vink}}]{Dimitriadis2014}
{Dimitriadis} G., {Chiotellis} A., {Vink} J., 2014, \mnras, 443, 1370

\bibitem[{{Dobrzycka} \& {Kenyon}(1994)}]{Dobrzycka1994}
{Dobrzycka} D., {Kenyon} S.~J., 1994, \aj, 108, 2259

\bibitem[{{Dobrzycka} {et~al}\mbox{.}(1996){Dobrzycka}, {Kenyon}, {Proga},
  {Mikolajewska}, \& {Wade}}]{Dobrzycka1996}
{Dobrzycka} D., {Kenyon} S.~J., {Proga} D., {Mikolajewska} J., {Wade} R.~A.,
  1996, \aj, 111, 2090

\bibitem[{{Drew} \& {Proga}(2000)}]{Drew2000}
{Drew} J.~E., {Proga} D., 2000, \nar, 44, 21

\bibitem[{{Edgar}(2004)}]{Edgar2004}
{Edgar} R., 2004, \nar, 48, 843

\bibitem[{{Evans} {et~al}\mbox{.}(2007){Evans}, {Woodward}, {Helton}, {van
  Loon}, {Barry}, {Bode}, {Davis}, {Drake}, {Eyres}, {Geballe}, {Gehrz},
  {Kerr}, {Krautter}, {Lynch}, {Ness}, {O'Brien}, {Osborne}, {Page}, {Rudy},
  {Russell}, {Schwarz}, {Starrfield}, \& {Tyne}}]{Evans2007}
{Evans} A. {et~al.}, 2007, \apjl, 671, L157

\bibitem[{{Fekel} {et~al}\mbox{.}(2000){Fekel}, {Joyce}, {Hinkle}, \&
  {Skrutskie}}]{Fekel2000}
{Fekel} F.~C., {Joyce} R.~R., {Hinkle} K.~H., {Skrutskie} M.~F., 2000, \aj,
  119, 1375

\bibitem[{{Ferguson} \& {Ferland}(1997)}]{Ferguson1997}
{Ferguson} J.~W., {Ferland} G.~J., 1997, \apj, 479, 363

\bibitem[{{Hachisu} {et~al}\mbox{.}(2006){Hachisu}, {Kato}, {Kiyota},
  {Kubotera}, {Maehara}, {Nakajima}, {Ishii}, {Kamada}, {Mizoguchi},
  {Nishiyama}, {Sumitomo}, {Tanaka}, {Yamanaka}, \& {Sadakane}}]{Hachisu2006}
{Hachisu} I. {et~al.}, 2006, \apjl, 651, L141

\bibitem[{{Hachisu}, {Kato} \& {Luna}(2007){Hachisu}, {Kato}, \&
  {Luna}}]{Hachisu2007}
{Hachisu} I., {Kato} M., {Luna} G.~J.~M., 2007, \apjl, 659, L153

\bibitem[{{Hamuy} {et~al}\mbox{.}(1995){Hamuy}, {Phillips}, {Maza}, {Suntzeff},
  {Schommer}, \& {Aviles}}]{Hamuy1995}
{Hamuy} M., {Phillips} M.~M., {Maza} J., {Suntzeff} N.~B., {Schommer} R.~A.,
  {Aviles} R., 1995, \aj, 109, 1

\bibitem[{{Han} \& {Podsiadlowski}(2004)}]{Han2004}
{Han} Z., {Podsiadlowski} P., 2004, \mnras, 350, 1301

\bibitem[{{Hernanz} \& {Jos{\'e}}(2008)}]{Hernanz2008}
{Hernanz} M., {Jos{\'e}} J., 2008, \nar, 52, 386

\bibitem[{{Hounsell} {et~al}\mbox{.}(2010){Hounsell}, {Bode}, {Hick},
  {Buffington}, {Jackson}, {Clover}, {Shafter}, {Darnley}, {Mawson}, {Steele},
  {Evans}, {Eyres}, \& {O'Brien}}]{Hounsell2010}
{Hounsell} R. {et~al.}, 2010, \apj, 724, 480

\bibitem[{{Iijima}(2008)}]{Iijima2008}
{Iijima} T., 2008, in Astronomical Society of the Pacific Conference Series,
  Vol. 401, RS Ophiuchi (2006) and the Recurrent Nova Phenomenon, {Evans} A.,
  {Bode} M.~F., {O'Brien} T.~J., {Darnley} M.~J., eds., p. 115

\bibitem[{{Kozma} \& {Fransson}(1992)}]{Kozma1992}
{Kozma} C., {Fransson} C., 1992, \apj, 390, 602

\bibitem[{{Kurosawa}, {Harries} \& {Symington}(2006){Kurosawa}, {Harries}, \&
  {Symington}}]{Kurosawa2006}
{Kurosawa} R., {Harries} T.~J., {Symington} N.~H., 2006, \mnras, 370, 580

\bibitem[{{Li} \& {van den Heuvel}(1997)}]{Li1997}
{Li} X.-D., {van den Heuvel} E.~P.~J., 1997, \aap, 322, L9

\bibitem[{{Maguire} {et~al}\mbox{.}(2013){Maguire}, {Sullivan}, {Patat},
  {Gal-Yam}, {Hook}, {Dhawan}, {Howell}, {Mazzali}, {Nugent}, {Pan},
  {Podsiadlowski}, {Simon}, {Sternberg}, {Valenti}, {Baltay}, {Bersier},
  {Blagorodnova}, {Chen}, {Ellman}, {Feindt}, {F{\"o}rster}, {Fraser},
  {Gonz{\'a}lez-Gait{\'a}n}, {Graham}, {Guti{\'e}rrez}, {Hachinger},
  {Hadjiyska}, {Inserra}, {Knapic}, {Laher}, {Leloudas}, {Margheim},
  {McKinnon}, {Molinaro}, {Morrell}, {Ofek}, {Rabinowitz}, {Rest}, {Sand},
  {Smareglia}, {Smartt}, {Taddia}, {Walker}, {Walton}, \&
  {Young}}]{Maguire2013}
{Maguire} K. {et~al.}, 2013, \mnras, 436, 222

\bibitem[{{Mastrodemos} \& {Morris}(1998)}]{Mastrodemos1998}
{Mastrodemos} N., {Morris} M., 1998, \apj, 497, 303

\bibitem[{{Mazzali} {et~al}\mbox{.}(2008){Mazzali}, {Sauer}, {Pastorello},
  {Benetti}, \& {Hillebrandt}}]{Mazzali2008}
{Mazzali} P.~A., {Sauer} D.~N., {Pastorello} A., {Benetti} S., {Hillebrandt}
  W., 2008, \mnras, 386, 1897

\bibitem[{{Mohamed}(2010)}]{Mohamed2010}
{Mohamed} S., 2010, PhD thesis, University of Oxford

\bibitem[{{Mohamed}, {Mackey} \& {Langer}(2012){Mohamed}, {Mackey}, \&
  {Langer}}]{Mohamed2012}
{Mohamed} S., {Mackey} J., {Langer} N., 2012, \aap, 541, A1

\bibitem[{{Mohamed} \& {Podsiadlowski}(2007)}]{Mohamed2007}
{Mohamed} S., {Podsiadlowski} P., 2007, in Astronomical Society of the Pacific
  Conference Series, Vol. 372, 15th European Workshop on White Dwarfs,
  {Napiwotzki} R., {Burleigh} M.~R., eds., p. 397

\bibitem[{{Mohamed} \& {Podsiadlowski}(2012)}]{Mohamed2012b}
{Mohamed} S., {Podsiadlowski} P., 2012, Baltic Astronomy, 21, 88

\bibitem[{{Moore} \& {Bildsten}(2012)}]{Moore2012}
{Moore} K., {Bildsten} L., 2012, \apj, 761, 182

\bibitem[{{Morrison} \& {McCammon}(1983)}]{Morrison1983}
{Morrison} R., {McCammon} D., 1983, \apj, 270, 119

\bibitem[{{Mukai}(2008)}]{Mukai2008}
{Mukai} K., 2008, in Astronomical Society of the Pacific Conference Series,
  Vol. 401, RS Ophiuchi (2006) and the Recurrent Nova Phenomenon, {Evans} A.,
  {Bode} M.~F., {O'Brien} T.~J., {Darnley} M.~J., eds., p.~84

\bibitem[{{Nelson} {et~al}\mbox{.}(2011){Nelson}, {Mukai}, {Orio}, {Luna}, \&
  {Sokoloski}}]{Nelson2011}
{Nelson} T., {Mukai} K., {Orio} M., {Luna} G.~J.~M., {Sokoloski} J.~L., 2011,
  \apj, 737, 7

\bibitem[{{O'Brien} {et~al}\mbox{.}(2006){O'Brien}, {Bode}, {Porcas}, {Muxlow},
  {Eyres}, {Beswick}, {Garrington}, {Davis}, \& {Evans}}]{O'Brien2006}
{O'Brien} T.~J. {et~al.}, 2006, \nat, 442, 279

\bibitem[{{Panagia} {et~al}\mbox{.}(2006){Panagia}, {Van Dyk}, {Weiler},
  {Sramek}, {Stockdale}, \& {Murata}}]{Panagia2006}
{Panagia} N., {Van Dyk} S.~D., {Weiler} K.~W., {Sramek} R.~A., {Stockdale}
  C.~J., {Murata} K.~P., 2006, \apj, 646, 369

\bibitem[{{Patat} {et~al}\mbox{.}(2007){Patat}, {Chandra}, {Chevalier},
  {Justham}, {Podsiadlowski}, {Wolf}, {Gal-Yam}, {Pasquini}, {Crawford},
  {Mazzali}, {Pauldrach}, {Nomoto}, {Benetti}, {Cappellaro}, {Elias-Rosa},
  {Hillebrandt}, {Leonard}, {Pastorello}, {Renzini}, {Sabbadin}, {Simon}, \&
  {Turatto}}]{Patat2007}
{Patat} F. {et~al.}, 2007, Science, 317, 924

\bibitem[{{Patat} {et~al}\mbox{.}(2011){Patat}, {Chugai}, {Podsiadlowski},
  {Mason}, {Melo}, \& {Pasquini}}]{Patat2011}
{Patat} F., {Chugai} N.~N., {Podsiadlowski} P., {Mason} E., {Melo} C.,
  {Pasquini} L., 2011, \aap, 530, A63

\bibitem[{{P{\'e}rez-Torres} {et~al}\mbox{.}(2014){P{\'e}rez-Torres},
  {Lundqvist}, {Beswick}, {Bj{\"o}rnsson}, {Muxlow}, {Paragi}, {Ryder},
  {Alberdi}, {Fransson}, {Marcaide}, {Mart{\'{\i}}-Vidal}, {Ros}, {Argo}, \&
  {Guirado}}]{Perez-Torres2014}
{P{\'e}rez-Torres} M.~A. {et~al.}, 2014, \apj, 792, 38

\bibitem[{{Pontzen} {et~al}\mbox{.}(2013){Pontzen}, {Ro{\v s}kar}, {Stinson},
  {Woods}, {Reed}, {Coles}, \& {Quinn}}]{pynbody}
{Pontzen} A., {Ro{\v s}kar} R., {Stinson} G.~S., {Woods} R., {Reed} D.~M.,
  {Coles} J., {Quinn} T.~R., 2013, {pynbody: Astrophysics Simulation Analysis
  for Python}. Astrophysics Source Code Library, ascl:1305.002

\bibitem[{{Price}(2007)}]{Price2007}
{Price} D.~J., 2007, \pasa, 24, 159

\bibitem[{{Proga} {et~al}\mbox{.}(2002){Proga}, {Kallman}, {Drew}, \&
  {Hartley}}]{Proga2002}
{Proga} D., {Kallman} T.~R., {Drew} J.~E., {Hartley} L.~E., 2002, \apj, 572,
  382

\bibitem[{{Raskin} \& {Kasen}(2013)}]{Raskin2013}
{Raskin} C., {Kasen} D., 2013, \apj, 772, 1

\bibitem[{{Rupen}, {Mioduszewski} \& {Sokoloski}(2008){Rupen}, {Mioduszewski},
  \& {Sokoloski}}]{Rupen2008}
{Rupen} M.~P., {Mioduszewski} A.~J., {Sokoloski} J.~L., 2008, \apj, 688, 559

\bibitem[{{Shen}, {Guillochon} \& {Foley}(2013){Shen}, {Guillochon}, \&
  {Foley}}]{Shen2013}
{Shen} K.~J., {Guillochon} J., {Foley} R.~J., 2013, \apjl, 770, L35

\bibitem[{{Shore} \& {Aufdenberg}(1993)}]{Shore1993}
{Shore} S.~N., {Aufdenberg} J.~P., 1993, \apj, 416, 355

\bibitem[{{Shore} {et~al}\mbox{.}(1996){Shore}, {Kenyon}, {Starrfield}, \&
  {Sonneborn}}]{Shore1996}
{Shore} S.~N., {Kenyon} S.~J., {Starrfield} S., {Sonneborn} G., 1996, \apj,
  456, 717

\bibitem[{{Simon} {et~al}\mbox{.}(2009){Simon}, {Gal-Yam}, {Gnat}, {Quimby},
  {Ganeshalingam}, {Silverman}, {Blondin}, {Li}, {Filippenko}, {Wheeler},
  {Kirshner}, {Patat}, {Nugent}, {Foley}, {Vogt}, {Butler}, {Peek},
  {Rosolowsky}, {Herczeg}, {Sauer}, \& {Mazzali}}]{Simon2009}
{Simon} J.~D. {et~al.}, 2009, \apj, 702, 1157

\bibitem[{{Sokoloski} {et~al}\mbox{.}(2006){Sokoloski}, {Luna}, {Mukai}, \&
  {Kenyon}}]{Sokoloski2006}
{Sokoloski} J.~L., {Luna} G.~J.~M., {Mukai} K., {Kenyon} S.~J., 2006, \nat,
  442, 276

\bibitem[{{Springel}(2005)}]{Springel2005}
{Springel} V., 2005, \mnras, 364, 1105

\bibitem[{{Stehle} {et~al}\mbox{.}(2005){Stehle}, {Mazzali}, {Benetti}, \&
  {Hillebrandt}}]{Stehle2005}
{Stehle} M., {Mazzali} P.~A., {Benetti} S., {Hillebrandt} W., 2005, \mnras,
  360, 1231

\bibitem[{{Sternberg} {et~al}\mbox{.}(2011){Sternberg}, {Gal-Yam}, {Simon},
  {Leonard}, {Quimby}, {Phillips}, {Morrell}, {Thompson}, {Ivans}, {Marshall},
  {Filippenko}, {Marcy}, {Bloom}, {Patat}, {Foley}, {Yong}, {Penprase},
  {Beeler}, {Allende Prieto}, \& {Stringfellow}}]{Sternberg2011}
{Sternberg} A. {et~al.}, 2011, Science, 333, 856

\bibitem[{{Stockdale} {et~al}\mbox{.}(2006){Stockdale}, {Kelley}, {Sramek},
  {van Dyk}, {Immler}, {Weiler}, {Williams}, \& {Panagia}}]{Stockdale2006}
{Stockdale} C.~J., {Kelley} M., {Sramek} R.~A., {van Dyk} S.~D., {Immler} S.,
  {Weiler} K.~W., {Williams} C.~L.~M., {Panagia} N., 2006, Central Bureau
  Electronic Telegrams, 396, 1

\bibitem[{{Storey} \& {Hummer}(1995)}]{Storey1995}
{Storey} P.~J., {Hummer} D.~G., 1995, \mnras, 272, 41

\bibitem[{{Stritzinger} {et~al}\mbox{.}(2010){Stritzinger}, {Burns},
  {Phillips}, {Folatelli}, {Krisciunas}, {Kattner}, {Persson}, {Boldt},
  {Campillay}, {Contreras}, {Krzeminski}, {Morrell}, {Salgado}, {Freedman},
  {Hamuy}, {Madore}, {Roth}, \& {Suntzeff}}]{Stritzinger2010}
{Stritzinger} M. {et~al.}, 2010, \aj, 140, 2036

\bibitem[{{Sullivan} {et~al}\mbox{.}(2006){Sullivan}, {Le Borgne}, {Pritchet},
  {Hodsman}, {Neill}, {Howell}, {Carlberg}, {Astier}, {Aubourg}, {Balam},
  {Basa}, {Conley}, {Fabbro}, {Fouchez}, {Guy}, {Hook}, {Pain},
  {Palanque-Delabrouille}, {Perrett}, {Regnault}, {Rich}, {Taillet}, {Baumont},
  {Bronder}, {Ellis}, {Filiol}, {Lusset}, {Perlmutter}, {Ripoche}, \&
  {Tao}}]{Sullivan2006}
{Sullivan} M. {et~al.}, 2006, \apj, 648, 868

\bibitem[{{Sutherland} \& {Dopita}(1993)}]{Sutherland1993}
{Sutherland} R.~S., {Dopita} M.~A., 1993, \apjs, 88, 253

\bibitem[{{Theuns} \& {Jorissen}(1993)}]{Theuns1993}
{Theuns} T., {Jorissen} A., 1993, \mnras, 265, 946

\bibitem[{{van den Heuvel} {et~al}\mbox{.}(1992){van den Heuvel},
  {Bhattacharya}, {Nomoto}, \& {Rappaport}}]{vandenHeuvel1992}
{van den Heuvel} E.~P.~J., {Bhattacharya} D., {Nomoto} K., {Rappaport} S.~A.,
  1992, \aap, 262, 97

\bibitem[{{Vaytet} {et~al}\mbox{.}(2011){Vaytet}, {O'Brien}, {Page}, {Bode},
  {Lloyd}, \& {Beardmore}}]{Vaytet2011}
{Vaytet} N.~M.~H., {O'Brien} T.~J., {Page} K.~L., {Bode} M.~F., {Lloyd} M.,
  {Beardmore} A.~P., 2011, \apj, 740, 5

\bibitem[{{Verner} \& {Ferland}(1996)}]{Verner1996a}
{Verner} D.~A., {Ferland} G.~J., 1996, \apjs, 103, 467

\bibitem[{{Verner} {et~al}\mbox{.}(1996){Verner}, {Ferland}, {Korista}, \&
  {Yakovlev}}]{Verner1996b}
{Verner} D.~A., {Ferland} G.~J., {Korista} K.~T., {Yakovlev} D.~G., 1996, \apj,
  465, 487

\bibitem[{{Walder}, {Folini} \& {Shore}(2008){Walder}, {Folini}, \&
  {Shore}}]{Walder2008}
{Walder} R., {Folini} D., {Shore} S.~N., 2008, \aap, 484, L9

\bibitem[{{Wheeler}, {Lecar} \& {McKee}(1975){Wheeler}, {Lecar}, \&
  {McKee}}]{Wheeler1975}
{Wheeler} J.~C., {Lecar} M., {McKee} C.~F., 1975, \apj, 200, 145

\bibitem[{{Williams} {et~al}\mbox{.}(2008){Williams}, {Mason}, {Della Valle},
  \& {Ederoclite}}]{Williams2008}
{Williams} R., {Mason} E., {Della Valle} M., {Ederoclite} A., 2008, \apj, 685,
  451

\bibitem[{{Worters} {et~al}\mbox{.}(2007){Worters}, {Eyres}, {Bromage}, \&
  {Osborne}}]{Worters2007}
{Worters} H.~L., {Eyres} S.~P.~S., {Bromage} G.~E., {Osborne} J.~P., 2007,
  \mnras, 379, 1557

\bibitem[{{Yaron} {et~al}\mbox{.}(2005){Yaron}, {Prialnik}, {Shara}, \&
  {Kovetz}}]{Yaron2005}
{Yaron} O., {Prialnik} D., {Shara} M.~M., {Kovetz} A., 2005, \apj, 623, 398

\bibitem[{{Zamanov} {et~al}\mbox{.}(2007){Zamanov}, {Bode}, {Melo}, {Bachev},
  {Gomboc}, {Stateva}, {Porter}, \& {Pritchard}}]{Zamanov2007}
{Zamanov} R.~K., {Bode} M.~F., {Melo} C.~H.~F., {Bachev} R., {Gomboc} A.,
  {Stateva} I.~K., {Porter} J.~M., {Pritchard} J., 2007, \mnras, 380, 1053

\bibitem[{{Zamanov} {et~al}\mbox{.}(2005){Zamanov}, {Bode}, {Tomov}, \&
  {Porter}}]{Zamanov2005}
{Zamanov} R.~K., {Bode} M.~F., {Tomov} N.~A., {Porter} J.~M., 2005, \mnras,
  363, L26

\end{thebibliography}
}

\label{lastpage}

\end{document}